\documentclass[aps,prd,showpacs,nofootinbib,floats,floatfix,preprintnumbers,onecolumn]{revtex4}

\usepackage{bm}
\usepackage{latexsym}
\usepackage{dcolumn}
\usepackage{amsfonts,amssymb,amsmath}
\usepackage{graphicx,epsfig}
\usepackage{psfrag}
\usepackage{multirow}

\def\beq{\begin{equation}}
\def\eeq{\end{equation}}
\def\bea{\begin{eqnarray}}
\def\eea{\end{eqnarray}}
\def\benu{\begin{enumerate}}
\def\eenu{\end{enumerate}}

\def\l{\left}
\def\r{\right}

\def\pa{\partial}
\def\f{\frac}
\def\p{\prime}

\begin{document}

\title{Perturbations in dark energy models with evolving speed of sound}
\author{Rizwan Ul Haq Ansari}\email[E-mail:~]{rizwan@iucaa.ernet.in}
\author{Sanil Unnikrishnan}\email[E-mail:~]{sanil@iucaa.ernet.in}
\affiliation{IUCAA, Post Bag 4, Ganeshkhind, Pune 411 007, India.}
\date{\today}
\begin{abstract}
The behavior of perturbation in scalar field dark energy and its
consequent effect on the cold dark matter (CDM) power spectrum is well understood to
be governed by the equation of state (EOS) parameter and the
effective speed of sound (ESS) of dark energy.
 In this paper, we investigate  whether dark energy
 models whose ESS are epoch dependent leaves any
 distinct imprints on the large scale CDM power spectrum.
 In particular, we compare the cases where the ESS is
 decreasing with time with those where it increases.
 The CDM power spectrum is found to be generically suppressed in
 these cases as compared to the $\Lambda$CDM model.
The degree of suppression at different length scales can, in
principle, reflect the evolving nature of the ESS of dark energy.
However, we find that the effect on the CDM power spectrum in
cases where the ESS of dark energy is evolving with constant EOS
parameter is significantly smaller as compared to the situation where
ESS is constant whereas EOS parameter is evolving.
Further, it is also shown that the effect of different evolution of
ESS for a given evolution of EOS parameter of dark energy on the CDM power spectrum
is significant only at the intermediate scales (around $k \sim 0.01h/Mpc$).
At scales much smaller and larger than the Hubble radius, it is the evolution of
EOS parameter of dark energy which governs the degree of suppression of CDM power spectrum
with respect to the $\Lambda$CDM model.

\end{abstract}
 \maketitle


\section{Introduction}

It is now evident from numerous cosmological observations  that
our Universe is undergoing accelerated expansion at the present
epoch \cite{Reiss-1998, Perlmutter-1998, Spergel-2007,
Komatsu-2008, Komatsu-2010}. This profound discovery triggered a
flood of investigations both at the theoretical and experimental
front to understand the reason for this accelerated expansion.
The theoretical explanation for
this observed late time accelerated expansion can broadly be
classified into two categories. The first explanation is based on
the assumption that gravity as described by the Einstein's general
theory of relativity (GTR) is valid on the cosmological scale.
This assumption then forces us to invoke some exotic matter known
as dark energy to drive the late time accelerated expansion of the
universe. If not the dark energy, then the only other way of
explaining this accelerated expansion is to modify the theory of gravity
from the standard  GTR. In the present work we shall only
consider the first case, viz., the one where the accelerated
expansion is driven by dark energy.

Numerous models of dark energy  have been investigated in the
literature, these include quintessence, tachyon, k-essence,
Chaplygin, gas etc.\ (for details, see
Refs.\cite{Paddy-2003,Sahni-2004,Copeland-2006,Sahni-2006} and the
references therein). It is now a big challenge to observationally
test the predictions of these models. One of the ways of
distinguishing various models of dark energy is to investigate how
cold dark matter (CDM) clusters on cosmological scales. In this
paper we shall investigate how a general class of scalar field
models of dark energy influences the large scale CDM power spectrum.

It is well known that scalar field models of dark energy  can lead
to degenerate evolution of the scale factor
\cite{Paddy-2002,Feinstein-2002}. Therefore, the evolution of
scale factor is insufficient to distinguish scalar field models of
dark energy. Hence, it is also necessary to take into account the
consequence of perturbation in dark energy on cosmological
observable such as CDM matter power spectrum. Although, the
evolution of scale factor and the growth of CDM perturbations does
not uniquely characterize
 the nature of scalar field lagrangian of the dark energy (see Ref.\cite{Sanil-2008b} for details),
  however, such studies can be useful in at least ruling out some of the models.

Evolution of perturbations in dark energy and how it influences
CDM power spectrum and the integrated Sachs-Wolfe (ISW) effect in
the cosmic microwave background (CMB)
 have been investigated in the literature \cite{Hu-2001, Erickson-2002, Weller-2003, DeDeo-2003, Bean-2004,
Gordon-2004, Gordon-2005, Corasaniti-2005, Hannestad-2005,
Abramo-2007, Sanil-2008a, Ballesteros-2008, Gorini-2008,
Park-2009, Dent-2009, Abramo-2009, Jassal-2009, Sapone-2009,
Sergijenko-2009, Jassal-2010, Putter-2010, Sanchez-2010,
Ballesteros-2010, Karwan-2011}. It is well understood from these
studies that CDM power spectrum is suppressed in a generic scalar
field model of dark energy as compared to the same in $\Lambda$CDM
model. The degree of suppression depends on how scale factor as
well as perturbations in dark energy evolves.
 It is important to take into account the role played by the
 perturbations in dark energy for the system of equations to be
 consistent with Eistein's equation
 \cite{Sanil-2008a,Park-2009}.

In general, any minimally coupled scalar field model of dark
energy can be characterized by the behavior of the following two
parameter: (1) equation of state (EOS) $w$, and (2) the effective
speed of sound (ESS) $c_{e}$.  Most studies on the influence of
perturbation in dark energy involves the case where $c_{e}$ and
$w$ are constants (see for instance,
Refs.\cite{Sapone-2009,Ballesteros-2010}). Perturbations in dark
energy models where ESS parameter $c_{e}$ is epoch dependent have
also been investigated \cite{DeDeo-2003,Putter-2010}. However, the
evolution of $c_{e}$ was fixed by the specific model of dark
energy under investigation. Our aim in this paper is to
parameterize the evolution of $c_{e}$ as a function of scale
factor and to  study how perturbations in such dark energy
models influences the CDM power spectrum.

In this paper, we shall consider a class of k-essence dark energy models
with lagrangian of the form $\mathcal{L}(X,\phi) = F(X) -
V(\phi)$, where  $X =(1/2)\partial_{\mu}\phi\partial^{\mu}\phi$ is
the kinetic term. We shall first present a method of
reconstructing the form of $F(X)$ and $V(\phi)$ from a given
evolution of EOS parameter $w(a)$ and  ESS parameter $c_{e}(a)$
with scale factor. Using this method, we shall reconstruct few
models of dark energy starting from a desired evolution of $w(a)$
and $c_{e}(a)$. The evolution of perturbation in dark energy is
investigated for the following cases (i) constant $w$ and $c_{e}$,
(ii) either of $w$ and $c_{e}$ is evolving, and (iii) both $w$ and
$c_{e}$ evolving. The growth of large scale CDM perturbation is
compared in these models. It is shown that in models where $w$ is
constant whereas $c_{e}$ is evolving,  the effect of different
functional form of $c_{e}(a)$ on the CDM power spectrum is
significantly smaller as compared to the case where $c_{e}$ is
constant and $w$ is evolving.

This paper is organized as follows. In the next section, we
discuss the necessary equations required for studying the
background evolution in a system of CDM and k-essence dark energy.
 Section III deals  with the evolution of perturbations in the longitudinal gauge
 in  a system consisting of CDM matter and scalar field dark energy.
 In section IV, we develop a method of reconstructing the Lagrangian density of the scalar field dark energy
 from given the evolution of equation of state
 parameter and sound speed of dark energy perturbations.
 The influence
 of dark energy perturbations on the CDM power spectrum in such reconstructed  models
 is studied in section V. 
 Finally in Sec.VI, we discuss in detail the effect of ESS parameter at different length scales in CDM power spectrum.
 Summary  and conclusions are given in
 section VI.

\section{Background Evolution}
We consider a spatially  flat homogenous  and isotropic
Friedmann-Robertson-Walker (FRW) universe with line element given
by
\begin{equation}
ds^{2} = dt^{2} - a^{2}(t)\l[dx^{2} + dy^{2} + dz^{2}\r],
\end{equation}
where $a(t)$ is the scale factor.
 The late time evolution of the universe is primarily governed by the properties
 of dark matter and dark energy. Hence, the late time evolution
 of the scale factor is determined by the following Friedmann equation
\begin{equation}
\l(\frac{\dot{a}}{a}\r)^2 = \l(\frac{8 \pi G }{3}\r)\,\left[
\rho_{m}(a) + \rho_{de}(a) \right],  \label{eqn::background eqn 1}
\end{equation}
where $\rho_{m}$ and $\rho_{d}$ are the energy densities of dark
matter and dark energy respectively.
If the dark matter and dark
energy are not coupled to each other, then each component
individually satisfies the following  conservation equation
\begin{equation}\label{conserve eqn}
  \dot{\rho} = -3H\l( \rho + p\r).
\end{equation}
Since the dark matter is non-relativistic ($p \simeq 0$), the
conservation equation [Eq.(\ref{conserve eqn})] implies that
\begin{equation}\label{matter conserve}
 \rho_{m} = \rho_{m_{0}}\,a^{-3},
\end{equation}
where $\rho_{m0}$ is the density of the dark matter at the present
epoch.
From the conservation equation  (Eq.(\ref{conserve eqn})),
it follows that the dark energy density evolves as
\begin{equation}
\rho_{de}(a) = \rho_{de_{0}}\exp\left(-3\int \l(\frac{1 +
w(a)}{a}\r)da\right), \label{DE conserve}
\end{equation}
where $\rho_{de0}$ is the dark energy density  at the present
epoch and $w(a)$ is the equation of state parameter of the dark
energy defined as
\begin{equation}\label{eos}
w(a)= \frac{p_{de}}{\rho_{de}}.
\end{equation}

For a given $w(a)$,  Eqs.(\ref{eqn::background eqn 1}), (\ref{matter
conserve}), and (\ref{DE conserve}) form a close set of equations
determining the evolution of scale factor. However, the evolution
of the equation of state parameter $w(a)$ with scale factor can
only be determined by the underlying nature of dark energy.

We assume  that the dark energy is a minimally coupled scalar
field with the action given by
\begin{equation}\label{action}
 S[\phi] =
\int\  \sqrt{-g}\; {\cal L}(X,\phi) \, d^{4}x,
\end{equation}
where ${\cal L}(X,\phi)$ is the lagrangian density of the scalar
field which is a function of scalar field $\phi$ and the kinetic
term $X$, which is defined as
\begin{eqnarray}
X =(1/2)\partial_{\mu}\phi\partial^{\mu}\phi.\nonumber
\end{eqnarray}
The stress energy-momentum tensor associated with the scalar
  field is given by
\begin{equation}\label{Scalar EM}
  T^{\mu\nu} = \l(\frac{\partial{\cal L}}{\partial X}\r)\, \l(\partial^{\mu}\phi\; \partial^{\nu}\phi\r) -
g^{\mu\nu}\, {\cal L}.
\end{equation}

 In the background space-time with the FRW line element, the
scalar field $\phi$ can only be a function of time $\phi=
\phi(t)$. Consequently the stress energy-momentum tensor will be
diagonal $T^{\mu\nu}= \text{diag}(\rho_\phi,- p_\phi, - p_\phi,-
p_\phi)$, with
\begin{eqnarray}
  \rho_{\phi} &=& \l(\f{\pa {\cal L}}{\pa X}\r)(2\, X)- {\cal L},\label{Scalar
  density}\\
  p_{\phi} &=& {\cal L}.\label{scalr pressure}
\end{eqnarray}

Therefore, the equation of state parameter of dark energy defined
in Eq.(\ref{eos})
 can be expressed  as
\begin{equation}\label{Eos scalar}
w=\frac{\mathcal{L}(X,\phi)}{2\l(\frac{\partial
\mathcal{L}}{\partial X}\r) X - \mathcal{L}(X, \phi)}.
\end{equation}
The above equation [Eq.(\ref{Eos scalar})] implies that the
evolution of EOS parameter $w(a)$ can be determined once the
evolution of the scalar field $\phi(t)$ on the FRW background is
known. The scalar field action (\ref{action}) implies that the
background
 field $\phi(t)$ satisfies the following equation of motion
\begin{equation}\label{EOM scalar}
\left[\l(\f{\pa {\cal L}}{\pa X}\r)+ 2\, X \l(\f{\pa^2 {\cal
L}}{\pa X^2}\r)\right]\ddot{\phi}+ \left[3H \l(\f{\pa {\cal
L}}{\pa X}\r)+ \dot{\phi} \f{\pa^2 {\cal L}}{\pa X \pa \phi
}\right]\dot{\phi}- \l(\f{\pa {\cal L}}{\pa \phi}\r)=0.
\end{equation}

 Given a Lagrangian density ${\cal L}(X,\phi)$, Eqs.(\ref{eqn::background eqn 1}) and  (\ref{EOM scalar})
 form a closed set of equations for  determining the evolution
$a(t)$ and $\phi(t)$. Consequently, the evolution of $w(a)$ can be
determined using Eq.(\ref{Eos scalar}). However, it is also
possible to reconstruct the form of Lagrangian density starting from from a
given evolution of the EOS parameter $w(a)$. We shall adopt this method
 of reconstruction in section III for designing few models of dark energy
 which leads to some desired evolution of $w(a)$.

\section{Evolution of scalar perturbation}

In order to study the evolution of perturbations in CDM and dark
energy we consider the following perturbed FRW metric in the
longitudinal gauge \cite{Bardeen-1980,Kodama-1984,Mukhanov-1992},
\begin{equation}
    ds^{2} = \left( 1 + 2\Phi\right)dt^{2} - a^{2}(t)\left(1 - 2 \Phi\right)\left[dx^2 + dy^2 +
    dz^2\right], \label{eqn::longitudinal gauge}
\end{equation}
where $\Phi$ is the variable describing  scalar metric
perturbations and it is also known as the Bardeen potential \cite{Bardeen-1980}.
The energy momentum tensor for the matter content of the universe such as CDM
and  scalar field dark energy can be expressed  as
\begin{equation}
T^{\mu}_{\hspace{0.2cm}\nu} = \left(\rho + p\right)u^{\mu}u_{\nu}
- p\delta^{\mu}_{\hspace{0.2cm}\nu}, \label{eqn A :: EM tensor}
\end{equation}
where $\rho$ is the energy density, $p$ is the pressure and
$u^{\mu}$ is the four velocity field. The perturbations in the
energy density, pressure and the four velocity field $u^{\mu}$ are defined in the following way
\begin{eqnarray}
\rho(t,\vec{x}) &=& \bar{\rho}(t) + \delta \rho(t,\vec{x}), \label{eqn A :: density perturbation}\\
p(t,\vec{x}) &=&  \bar{p}(t) + \delta p(t,\vec{x}), \label{eqn A :: pressure perturbation}\\
u^{\mu} &=& \bar{u}^{\mu} + \delta u^{\mu}, \label{eqn A ::
velocity perturbation}
\end{eqnarray}
where $\bar{\rho}(t)$ and $\bar{p}(t)$ are the energy density and pressure
in the background FRW line element, respectively, and $\bar{u}^{\mu} = [1,0,0,0]$
is the background four velocity. Since $u^{\mu}u_{\mu} = 1$, it
turns out that $\delta u_{0} = - \delta u^{0} = \Phi $. Further,
the spatial part of  the four velocity field $\delta u^{i}$  can be expressed as a gradient of a scalar
\begin{eqnarray}\label{velocity perturbation 1}
\delta u^{i} = \delta^{ij}u_{\hspace{0.05cm},\hspace{0.05cm}j}.
\end{eqnarray}

From Eqs.(\ref{eqn A :: EM tensor}) to (\ref{velocity
perturbation 1}), it follows that the perturbations in the energy
momentum tensor can be written in terms of $\delta \rho$, $\delta
p$ and $u$ as
\begin{eqnarray*}
  \delta T^{0}_{\hspace{0.2cm}0} &=& \delta \rho, \\
  \delta T^{0}_{\hspace{0.2cm}i} &=& \left(\bar{\rho} + \bar{p}\right)u_{,i} \\
 \delta T^{i}_{\hspace{0.2cm}j} &=& -\delta p\, \delta^i_{~j}.
\end{eqnarray*}
The three variables  $\delta \rho$, $\delta p$ and $u$ describes
the scalar degree of perturbations in the matter sector for both
perfect fluid and scalar field. In the case of pressureless
matter ($\bar{p}=\delta p_m =0$), the perturbations are described
by $\delta \rho_m$ and $u_m$. For the scalar field dark energy,
the perturbation in the scalar field $\phi$ is defined as
\begin{equation}\label{scalar field pert}
    \phi(\vec{x},t) = \bar{\phi}(t) + \delta\phi(\vec{x},t),
\end{equation}
where $\bar{\phi(t)}$ is the value of scalar field on the
background FRW spacetime.
Substituting Eq.(\ref{scalar field pert})
in the stress-energy momentum tensor of scalar field tensor
defined in Eq.(\ref{Scalar EM}) and subtracting the background $\rho_{\phi}$
 and $p_{\phi}$ we get
\begin{eqnarray}
\delta\rho_{\phi} &=& \left(\dot{\bar{\phi}}\dot{\delta\phi} - \Phi \dot{\bar{\phi}}^2\right)\left[\l(\frac{\partial\mathcal{L}}{\partial X}\r) + \l(2 X\r)\l(\frac{\partial^{2}\mathcal{L}}{\partial X^{2}}\r)\right]
 -  \left[\l(\frac{\partial \mathcal{L}}{\partial \phi}\r)-  \l(2X\r)\l(\frac{\partial^{2}\mathcal{L}}{\partial X\partial \phi}\r)\right]\delta\phi, \label{eqn A :: perturbed density NC scalar field}\\
\delta p_{\phi} &=& \left(\dot{\bar{\phi}}\dot{\delta\phi} - \Phi \dot{\bar{\phi}}^2\right)\l(\frac{\partial \mathcal{L}}{\partial X}\r) +   \l(\frac{\partial \mathcal{L}}{\partial \phi}\r)\delta\phi, \label{eqn A :: perturbed pressure NC scalar field}\\
u_{\phi} &=& -\l(\frac{\delta \phi}{a^{2}\dot{\bar{\phi}}}\r). \label{eqn A ::
peculiar velocity NC scalar field}
\end{eqnarray}

In the system consisting  of CDM and dark energy, the linearized Einstein's
equation $\delta G^\mu_{ \hspace{0.2cm} \nu} = 8\pi G\,\delta T^\mu_{\hspace{0.2cm} \nu}$, which relates the
matter perturbations to the metric perturbations, leads to the following equations
\begin{eqnarray}
3H^{2}\Phi + 3 H\dot{\Phi} + \frac{k^{2}}{a^{2}} \Phi  &=& - 4 \pi G\, \l( \bar{\rho}_m(a)\delta_m + \bar{\rho}_d(a)\delta_d\r), \label{Lin Ein1} \\
\dot{\Phi} + H \Phi &=& - 4 \pi G a^{2}\,\l(\bar{\rho}_m(a) u_m +  (\bar{\rho}_d + \bar{p_d})u_d\r), \label{Lin Ein2} \\
\ddot{\Phi} + 4H\dot{\Phi} + \left(2\dot{H} + 3H^{2}\right)\Phi
&=& 4 \pi G\, \delta p_{d}, \label{Lin Ein3}
\end{eqnarray}
where  $\delta_{m}$ and $\delta_{d}$ are the fractional density perturbation\footnote{Hereafter, we will be using the subscript `d' instead of $\phi$ in the variables describing perturbations in dark energy} in CDM matter and dark energy, respectively,  defined as
\begin{eqnarray*}
\delta_{m} &\equiv& \frac{\delta\rho_{m}}{\bar{\rho}_m}\\
\delta_{d} &\equiv& \frac{\delta\rho_{d}}{\bar{\rho}_{d}}.
\end{eqnarray*}
It should be understood from the above Einstein's equations [Eqs.(\ref{Lin Ein1})-(\ref{Lin Ein3})] that each of the
variables describing the perturbations such as
$\Phi,\,\delta\rho_{m}, \,\delta\rho_{d}$, etc.\ are in fact the
amplitude in the fourier space for a given fourier mode $k$.

The covariant conservation equation $T^{\mu} \hspace{0.03cm}_{\nu\, ;\,\mu }=0$ implies that for the
pressureless matter such as CDM
\begin{eqnarray*}
\dot{\delta}_{m} &=&  k^{2}u_{m} + 3\dot{\Phi},\\\label{eqn :: PE1}
\dot{u}_{m} &=& -2Hu_{m} - \frac{\Phi}{a^{2}},\label{eqn :: PE2}
\end{eqnarray*}
where overdot denotes derivative with respect to cosmic time.
In the case of scalar field dark energy the perturbations in pressure
$\delta p_{d}$ is related to $\delta\rho_{d}$ and $u_d$ as \cite{Hu-1998, Sanil-2008b, Sanil-2010}
\begin{eqnarray}
\delta p_{d} = c_{e}^{2} \delta \rho_{d} -3H\left(\bar{\rho}_{d} +
\bar{p}_{d}\right)a^{2}u_{d}\left[c_{e}^{2} -
c_{a}^{2}\right], \label{eqn  :: delta p rho}
\end{eqnarray}
where $c_{a}^{2}$ is the square of the adiabatic  speed of sound defined as
\begin{eqnarray}
c_{a}^{2} \equiv \frac{\dot{\bar{p}}_{d}}{\dot{\bar{\rho}}_{d}} = w -
\frac{\dot{w}}{3H\left(1 + w\right)}, \label{PE 3}
\end{eqnarray}
and $c_{e}^{2}$ is square of the effective speed of sound of scalar field given by \cite{Garriga-1999,Picon-1999}
\begin{equation}
c_{e}^{2} =
\frac{\l(\frac{\partial \mathcal{L}}{\partial X}\r)}{\l(\frac{\partial \mathcal{L}}{\partial X}\r)
+ 2
 X\l(\frac{\partial^{2}\mathcal{L}}{\partial X^{2}}\r)}.\label{eqn
: effective sound speed dark energy}
\end{equation}
The ESS parameter $c_{e}^{2}$ is in fact the ratio of dark energy pressure perturbation to the energy density perturbation in the comoving gauge or rest frame of dark energy \cite{Gordon-2004,Hu-2005}.
In the case of scalar field dark energy this gauge coincides with the uniform field gauge.

For the dark energy density perturbation, the covariant conservation equation ($T^{\mu} \hspace{0.03cm}_{\nu\, ;\,\mu }=0$) together with Eq.(\ref{eqn  :: delta p rho}) leads to the following equations
\begin{eqnarray}
\dot{\delta}_{d} &=&  \left(1 + w\right)k^{2}u_{d} +  3H\left(w - c_{e}^{2}\right)\delta_{d} + 9H^{2}
  \left( 1 + w\right)\left[c_{e}^{2} - c_{a}^{2}\right]a^{2}u_{d} +  3\left(1 + w\right)\dot{\Phi},\label{eqn :: PE3} \\
 \dot{u}_{d} &=& -H\left(2 - 3c_{e}^{2}\right)u_{d} - \frac{c_{e}^{2}\delta_{d}}{a^{2}\left(1 + w\right)}    -  \frac{\Phi}{a^{2}}.
 \label{eqn :: PE4}
\end{eqnarray}
We now introduce the following dimensionless variable $\sigma_d$ defined as
\begin{equation}
\sigma_d \equiv 3 H \l(1+w\r) a^2 u_d
\end{equation}

For studying the evolution of perturbations in a system of CDM matter and dark energy,
it is evident that Eqs.(\ref{Lin Ein3}), (\ref{eqn :: PE3}) and (\ref{eqn :: PE4}) forms a closed set of equations.
In terms of the variables $\Phi$, $\delta_{d}$  and $\sigma_d$, these
equations can be re-expressed as
 \begin{eqnarray}
\Phi^{\p\p} &=& - \frac{\Phi^{\p }}{a}\left( \frac{\ddot{a}}{aH^2}+4
\right)- \frac{\Phi}{a^2}\left( 2\frac{\ddot{a}}{aH^2}+ 1 \right)
+ \frac{4 \pi G \bar{\rho}_{d} }{a^2 H^2}\l( c_{e}^{2} \delta_{d}
 -3H\left(1 + w \right)a^{2}u_{d}\left[c_{e}^{2} -
c_{a}^{2}\right]\r) \label{Phi prime}\\
  \sigma^{\p}_{d} &=& \left( \l(\frac{\ddot{a}}{aH^2}\r) +3\left(c_{e}^{2} -
c_{a}^{2}\right)+ \l(3w -1\r)\right)\frac{\sigma_{d}}{a}-\frac{3c_{e}^{2}\delta_{d}}{a}-\frac{3\Phi
(1+w)}{a}\label{sigma}\\
 \delta^{\p}_{d} &=&  \l(\frac{3\, \delta_{d}}{a}\r)\left(w - c_{e}^{2}\right) +
\frac{3\,\sigma_{d}}{a}\left(c_{e}^{2} - c_{a}^{2}\right)+\frac{k^2\,\Phi}{3 a^3 H^2}
 + 3\left(1 + w\right)\Phi^{\p}\label{delta d}
 \end{eqnarray}
where prime denotes the derivative with respect to the scale factor $a$.
Once the solutions $\Phi(a,\,k)$ and $\delta_{d}(a,\, k)$ are determined from the above three equations,
the evolution of  perturbation in CDM can be evaluated from the time-time
component of the Einstein's equation and this is given by
\begin{eqnarray}\label{CDM pert}
 \delta_{m}(a,\, k)= -\l(\frac{2}{\Omega_{m}(a)}\r)\l(\Phi + a\, \Phi^{\p}+ \l(\frac{k^2 }{3H^{2}a^2}\r)\Phi\r)-\l(\frac{\Omega_{d}(a)}{\Omega_{m}(a)}\r)\,\delta_{d}(a,\, k),
\end{eqnarray}
where $\Omega_{m}$ and $\Omega_{d}$ are the dimensionless density parameter, defined as
\begin{eqnarray*}
\Omega_{m}(a) &\equiv& \l(\frac{8\pi G }{3H^{2}}\r)\bar{\rho}_{m}(a)\\
\Omega_{d}(a) &\equiv& \l(\frac{8\pi G }{3H^{2}}\r)\bar{\rho}_{d}(a)
\end{eqnarray*}

\section{Reconstructing dark energy models}

It is evident from Eqs.(\ref{Phi prime})-(\ref{delta d}) that the
evolution of perturbations in the matter and dark energy are
governed by the behavior of dark energy speed of sound $c_e^2(a)$
and equation of state parameter $w(a)$. As mentioned earlier the
evolution of $c_e^2(a)$ and $w(a)$, in general, depends on the
underlying nature of the lagrangian density of the scalar field
describing the dark energy. However, one could also reconstruct
the form of the lagrangian density of the scalar field dark energy
from a given evolution of ESS and EOS parameter. In this section
we will present such a method of reconstruction. The evolution of
perturbation in such reconstructed models will be discussed in the
following section.

In this paper we restrict our analysis to the following class of
scalar field dark energy models with the lagrangian density given
by \cite{Mukhanov-2006}:
\begin{equation}\label{Recons lagrang }
  \mathcal{L}(X,\phi) = F(X)-V(\phi).
\end{equation}
This is a natural generalization of the quintessence model where
$F(X)=X$. The form of the kinetic function $F(X)$ and potential
$V(\phi)$ can be reconstructed from a given evolution of $w(a)$
and $c_e^2(a)$. It is important to note that such class of dark
energy models allows evolution in $c_e^2(a)$ even when $w$ is
constant. This is unlike the tachyon model ($\mathcal{L}(X,\phi) =
-V(\phi)\sqrt{1 - 2X}$) where $c_e^2(a) = - w(a)$ and
consequently, a constant equation of state always necessarily
implies that $c_e^2$ is also constant.

For the model described by the lagrangian density (\ref{Recons
lagrang }), the field equation for the $\phi$ given by
Eq.(\ref{EOM scalar}) becomes
\begin{equation}\label{Field eqn}
  \ddot{\phi}+ 3H \dot{\phi} \l(\frac{F_X}{F_X+2XF_{XX}}\r) +
  \frac{V_\phi}{F_X+2XF_{XX}}=0.
\end{equation}
The ESS parameter $c_{e}^{2}$ for this model reads
\begin{equation}\label{Ess recons}
c_{e}^{2} = \frac{F_X}{F_X + 2 X F_{XX}}
\end{equation}
 Further, the EOS
parameter $w(a)$ can be expressed as
\begin{equation}\label{Eos recons}
 w=\frac{\mathcal{L}}{2 X F_{XX}-\mathcal{L}}.
\end{equation}
Using Eqs.(\ref{Ess recons}) and (\ref{Eos recons}), the equation
of motion for the scalar field $\phi$ can be re-expressed as
\begin{equation}\label{X dot}
  \frac{\dot{X}}{X}= -c_e^2 \l[ 6H + \l( \frac{2}{1+w}\r)\frac{\dot{V}}{\bar{\rho}_d(a)}
  \r],
\end{equation}
where $X= \frac{1}{2} \dot{\phi^2}$ is the kinetic term for the
background field  $\phi$.

From the definition of adiabatic speed of sound (see
Eq.(\ref{PE 3}), it follows that
\begin{equation}\label{Adiab
speed}
  c_a^2=\frac{\dot{V}-F_X \dot{X}}{6 H X F_X},
\end{equation}
which on rearranging implies that
\begin{equation}\label{X dot1}
  \frac{\dot{X}}{X}= \l(
  \frac{2}{1+w}\r)\l(\frac{\dot{V}}{\bar{\rho}_d(a)}\r)-6H c_a^2.
\end{equation}

The two equations [Eq.(\ref{X dot}) and (\ref{X dot1})] are the
evolution equations for the scalar field, with the first one being
the field equation derived from the scalar field action and second
one follows from the definition of the adiabatic speed of sound.
Equating the two equations we get
\begin{equation}\label{DV}
  \frac{dV}{da}= -3\rho_{d0}\l[
  \frac{c_e^2-c_a^2}{1+c_e^2}\r]\l(\frac{1+w(a)}{a}
  \r)\hbox{exp}\l[-3\int\l(\frac{1+w(a)}{a}\r)da  \r]
\end{equation}
\begin{equation}\label{X}
X(a)= X_0 \hbox{exp}\l[-6\int c_e^2 \l[\frac{1+c_a^2}{1+c_e^2}\r]
\frac{da}{a} \r]
\end{equation}
where $X_0$ is the constant of integration. Since, $X= \frac{1}{2}
\dot{\phi^2}$, the evolution of scalar field with the scale factor
is given by
\begin{equation}\label{Phi}
\phi(a)=\int \frac{\sqrt{2X(a)}}{aH(a)} da.
\end{equation}

Given any functional form of $c_e^2(a)$ and $w(a)$, Eq.(\ref{X})
can be integrated to obtain $X(a)$. Consequently, the evolution of
scalar field with the scale factor $\phi(a)$ can be determined
from Eq.(\ref{Phi}). Further, integration of Eq.(\ref{DV}) would
give potential $V(a)$ as a function of scale factor. By
eliminating the scale factor `$a$' from the two functions $V(a)$
and $\phi(a)$, the form of the potential $V(\phi)$ can be
determined.

The form of the kinetic function $F(X)$ can be reconstructed in
the following way. From the definition of the EOS parameter (see
Eq.(\ref{Eos recons})), it follows that
\begin{equation}\label{recons1}
  \frac{dF}{d\ln X}=\l( \frac{ 1+ w(a)}{2}\r) \rho_d(a).
\end{equation}
Using Eq.(\ref{X}), the above equation can be re-expressed as
\begin{equation}\label{recons4}
  \frac{dF}{da}= -c_e^2 \l[
  \frac{1+c_a^2}{1+c_e^2}\r]\l( \frac{3(1+w)}{a}
  \r)\rho_{d0}\, \hbox{exp}\l[-3\int\l(\frac{1+w(a)}{a}da \r) \r].
\end{equation}
From the solution of the above equation, it is possible to
determine how the kinetic function $F(a)$ evolves with scale
factor, for a given evolution of $w(a)$ and $c_e^2$. Further, from
the two solutions: $X(a)$ obtained in Eq.(\ref{X}) and
$F(a)$ from Eq.(\ref{recons4}), one can eliminate the scale factor to reconstruct the form
of the kinetic function $F(X)$.
It may not always be possible to reconstruct the exact analytical form of $F(X)$ and $V(\phi)$,
however, one can always numerically determine the functional form of these functions.
As our primary interest in this paper is to investigate how dark energy perturbations influences the CDM power spectrum, we will not be discussing the numerical form of $F(X)$ and $V(\phi)$ in each and every case.
The main purpose of this section was to emphasize the fact that the Lagrangian density of the form
$\mathcal{L}(X,\phi) = F(X)-V(\phi)$ indeed allows a class of solutions with evolution in $c_e^2(a)$ and $w(a)$.
\begin{figure}[t]
{\includegraphics[scale=0.40]{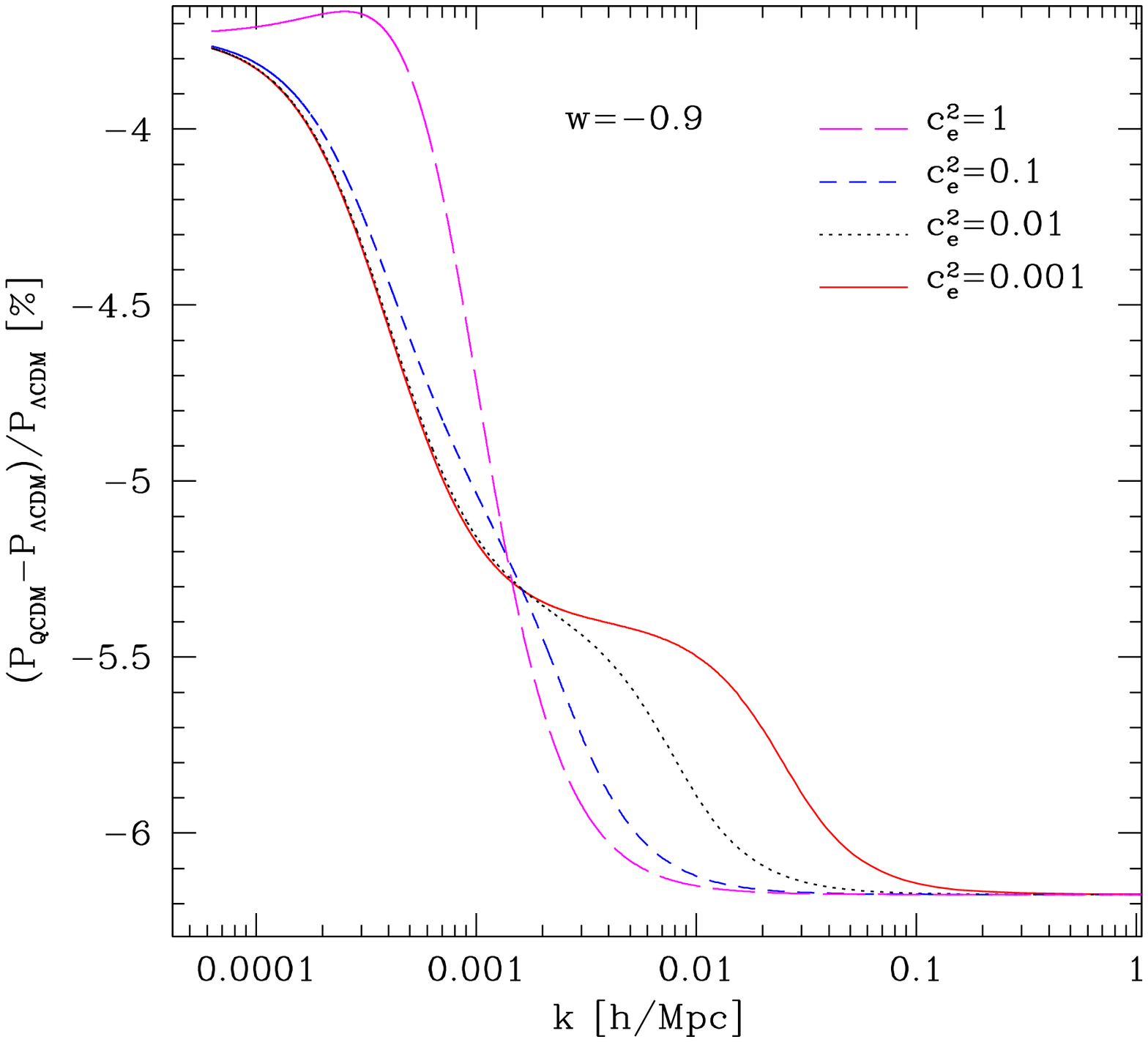}}
{\includegraphics[scale=0.40]{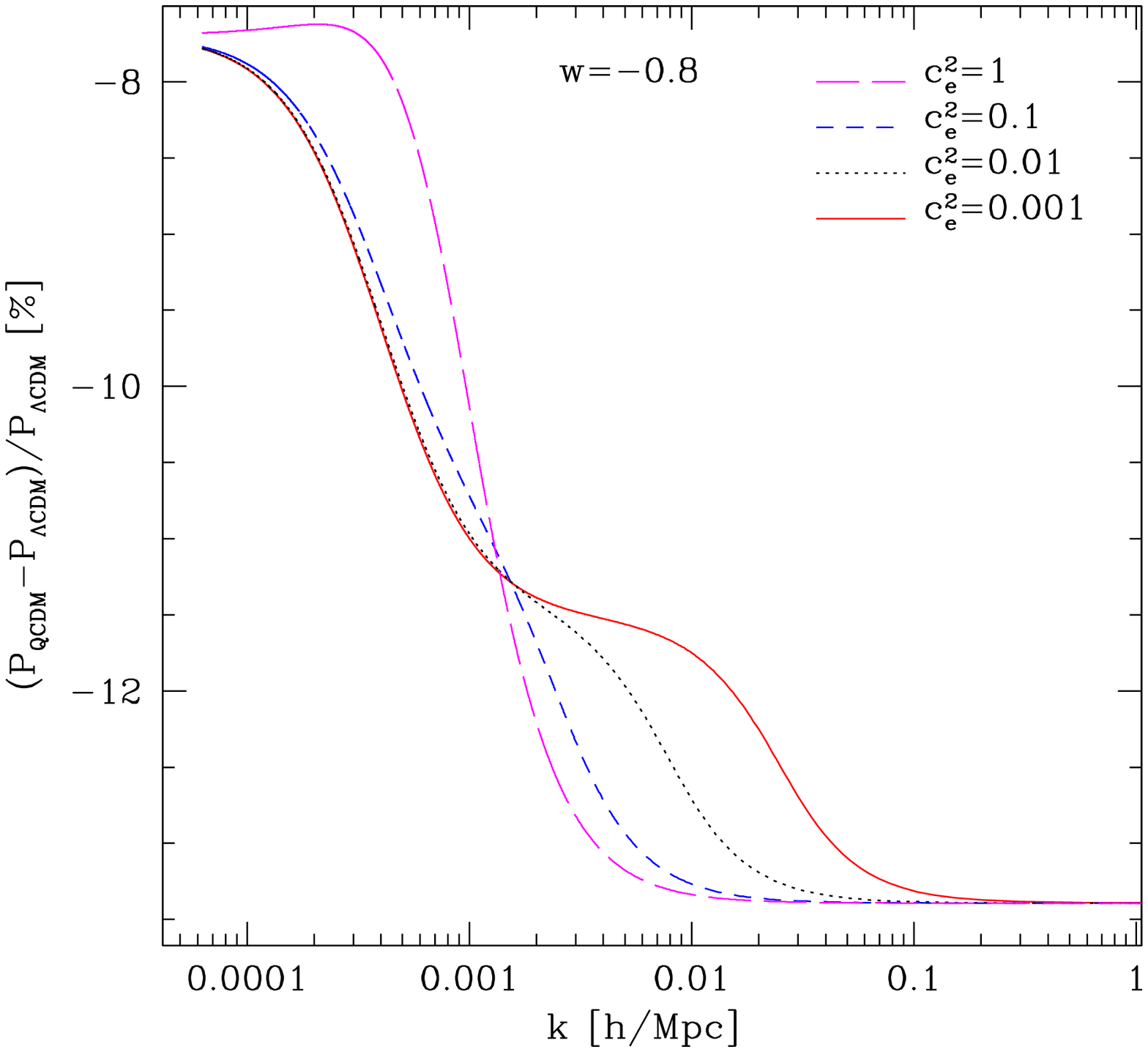}}
\caption{The percentage suppression
in matter power spectrum in dark energy models in comparison to $\Lambda$CDM model is plotted for a constant equation of state $w$ and for
different values of $c_e^2$.
The  left panel is for $w=-0.9$ and right
panel shows suppression for $w=-0.8$. It can be seen that
suppression is more for dark energy  models with  $w=-0.8$ with respect to $w=-0.9$.
The power spectrums shown in this figure corresponds to its value at the present epoch.}\label{F1}
\end{figure}

\section{Suppression of CDM power spectrum}
Our aim in this paper is to investigate  how perturbations in the
dark energy  influence the CDM matter power spectrum.
In particular, we are interested in models of dark energy whose speed of sound is epoch dependent.
In the preceding section, it was illustrated that it is possible to reconstruct the Lagrangian density of the form
$\mathcal{L}(X,\phi) = F(X)-V(\phi)$ from a given evolution of the EOS and ESS parameters.
Hence, in principle, it is always possible to associate $F(X)$ and $V(\phi)$ for a given evolution of $c_e^2(a)$ and $w(a)$.
This allowed us to choose a parameterized form of $c_e^2(a)$ and $w(a)$ for which the evolution of perturbations can be investigated.
But, before proceeding to the case where $c_e^2(a)$ is epoch dependent, we will first discuss the case where it is constant.

\begin{figure}
{\includegraphics[scale=0.40]{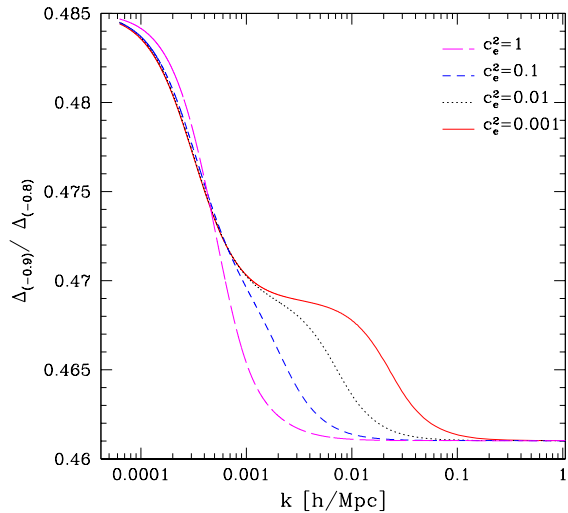}} \caption{Plot shows
the ratio of the percentage suppression in matter power spectrum $\Delta$ defined in Eq. (\ref{Delta}) for  $w=-0.9$
to $w=-0.8$, for different values of effective speed of
sound.}\label{F2}
\end{figure}

\subsection*{Case I: Constant $c_e^2$ and $w$ }
Considering the class of scalar field dark energy models with Lagrangian density of the form
$\mathcal{L}(X,\phi) = F(X)-V(\phi)$, if both $w$ and $c_e^2(a)$ are constants, it follows from Eqs.(\ref{X}) and (\ref{recons4}) that $X(a) \propto a^{n_{1}}$ and $F(a) \propto a^{n_{2}}$, where $n_{1}$ and $n_{2}$ are constants.
Therefore, when both $w$ and $c_e^2(a)$ are constants, the functional form of the kinetic function $F(X)$ would become
\cite{Mukhanov-2006}
$$F(X)=X^\alpha$$
where $\alpha$ is a constant given by
$$\alpha=\frac{1-c_e^2}{2c_e^2}.$$
It should be noted that for models with $F(X)=X^\alpha$, the ESS parameter $c_e^2$  is always constant irrespective of the functional form of the potential $V(\phi)$ and $w(a)$.
It is the form of the potential which would lead to a constant value for the equation of state parameter.
The form of the potential can be numerically determined from Eqs.(\ref{DV}) and (\ref{Phi}).

In order to determine how CDM perturbation grows when both $c_e^2$ and $w$ are constants,
we numerically solve the
closed set of equations (\ref{Phi prime}) to (\ref{delta d}), which takes into account the role
played by the perturbations in dark energy.
Once the evolution of
$\Phi(a)$, $\delta_d(a)$ and $\sigma_d(a)$ is known, the growth of
CDM perturbations can be determined
 using equation (\ref{CDM pert}).
For quantifying the difference in the growth of CDM perturbation in dark energy models with respect to $\Lambda$CDM model,
we numerically evaluate the following quantity

\begin{equation}\label{Delta}
\Delta \% \equiv \frac{P_{_{QCDM}}(k) - P_{_{\Lambda CDM}}(k)}{P_{_{\Lambda CDM}}(k)}\times 100
\end{equation}
where $P_{_{QCDM}}(k)$ is the CDM power spectrum at the present epoch in dark energy models whereas
$P_{_{\Lambda CDM}}(k)$ is the corresponding power spectrum in $\Lambda$CDM model.
This quantity for the case of constant $c_e^2$ and $w$ is plotted in Fig. \ref{F1}.
It is clear from this figure that $\Delta$ is negative at all length scales of perturbations.
Therefore, CDM power spectrum is suppressed in dark energy models as compared to the same in $\Lambda$CDM model.

The degree of suppression in  CDM power spectrum depends on the value of $w$ and $c_e^2$ and on the length scale of perturbation.
In Fig. \ref{F1}, we have plotted the percentage suppression $\Delta \%$ for $w = -0.9$ and $w = - 0.8$ and for different value of constant $c_{e}^{2}$.
It follows from this figure that there is more suppression if $w = -0.8$ than if $w = -0.9$,
indicating the fact that the degree of suppression is in a way  proportional to the deviation of $w$ from $-1$.
When comparing  percentage suppression for $w = -0.8$ and $w = -0.9$ in Fig. \ref{F1}, it may appear that they differs just by a factor which is independent of $c_{e}^{2}$.
However, this is not true as illustrated by Fig. \ref{F2}, where we have plotted the ratio of $\Delta$ (defined in Eq.(\ref{Delta})) for $w = -0.8$ and $w = -0.9$.
This ratio does depend on the value of $c_{e}^{2}$ clarifying the fact that the percentage suppression $\Delta$ is not merely a product of two functions with one depending on $w$ and the other on $c_{e}^{2}$.

Further, for any value of $w$ and $c_e^2$, the percentage suppression decreases with increasing length scale of perturbations.
This statement is true for the case of  constant $w$, however,  in general, it depends on the functional form of the equation of state parameter $w(a)$.

It can also be observed from Fig. \ref{F1} that the effect of different values of $c_e^2$ (assumed to be constant) for a given value of $w$ on the CDM power spectrum is insignificant on scales much smaller than the Hubble radius.
This statement is also true for scales much larger than the Hubble radius.
In fact, it can be be seen from Fig. \ref{F1} that only at the intermediate scales, say at $k = 0.01 h/Mpc$, the effect of different $c_e^2$ is more pronounced.
At these scales, it turns out that the percentage suppression increases with the increase in the value of $c_e^2$.
The reason for the fact that the effect of different values of $c_e^2$ is significant only at the intermediate scales will be discussed in detail in the next section (Sec.VI).

\begin{figure}
\includegraphics[scale=0.40]{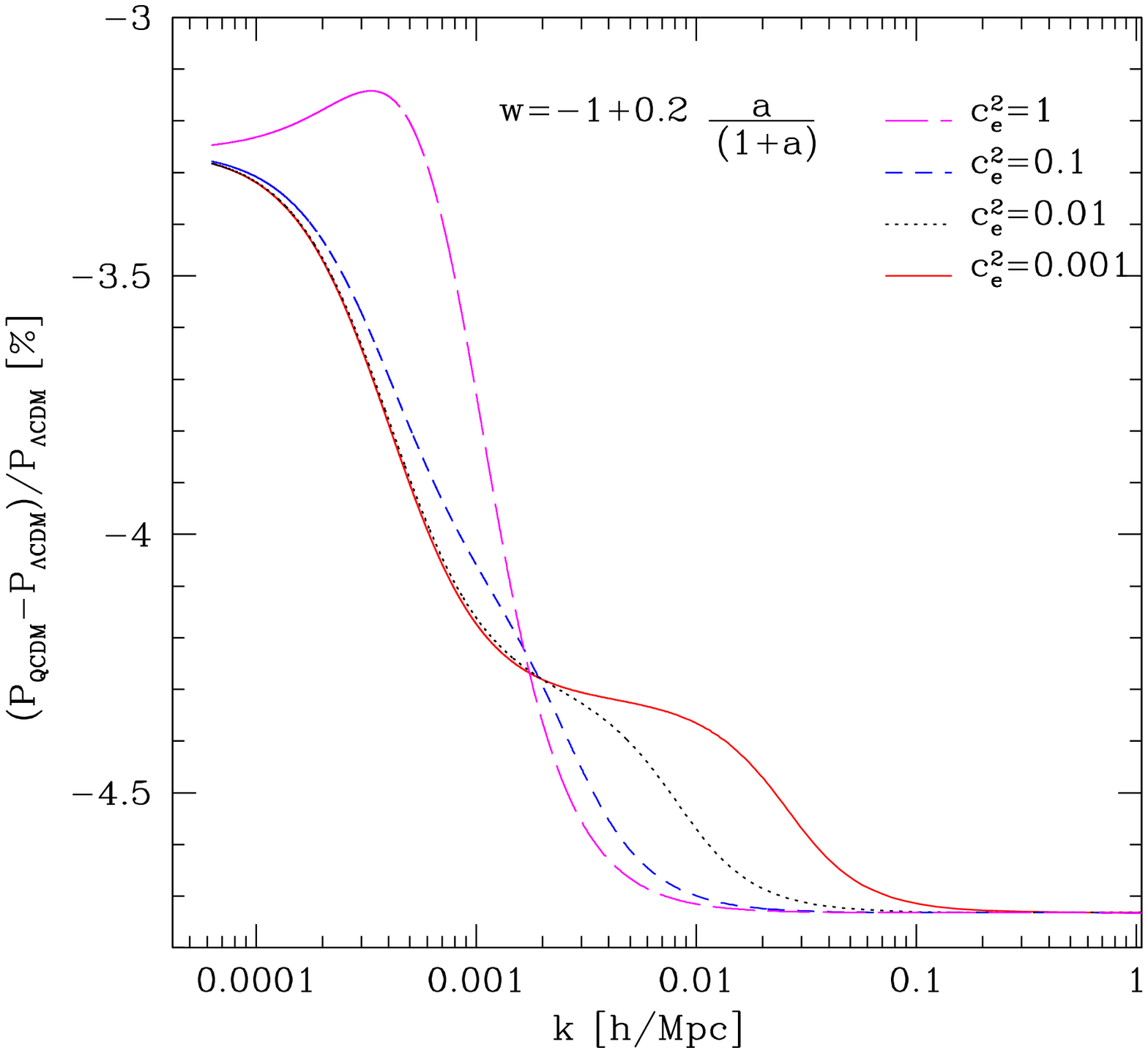}
\includegraphics[scale=0.40]{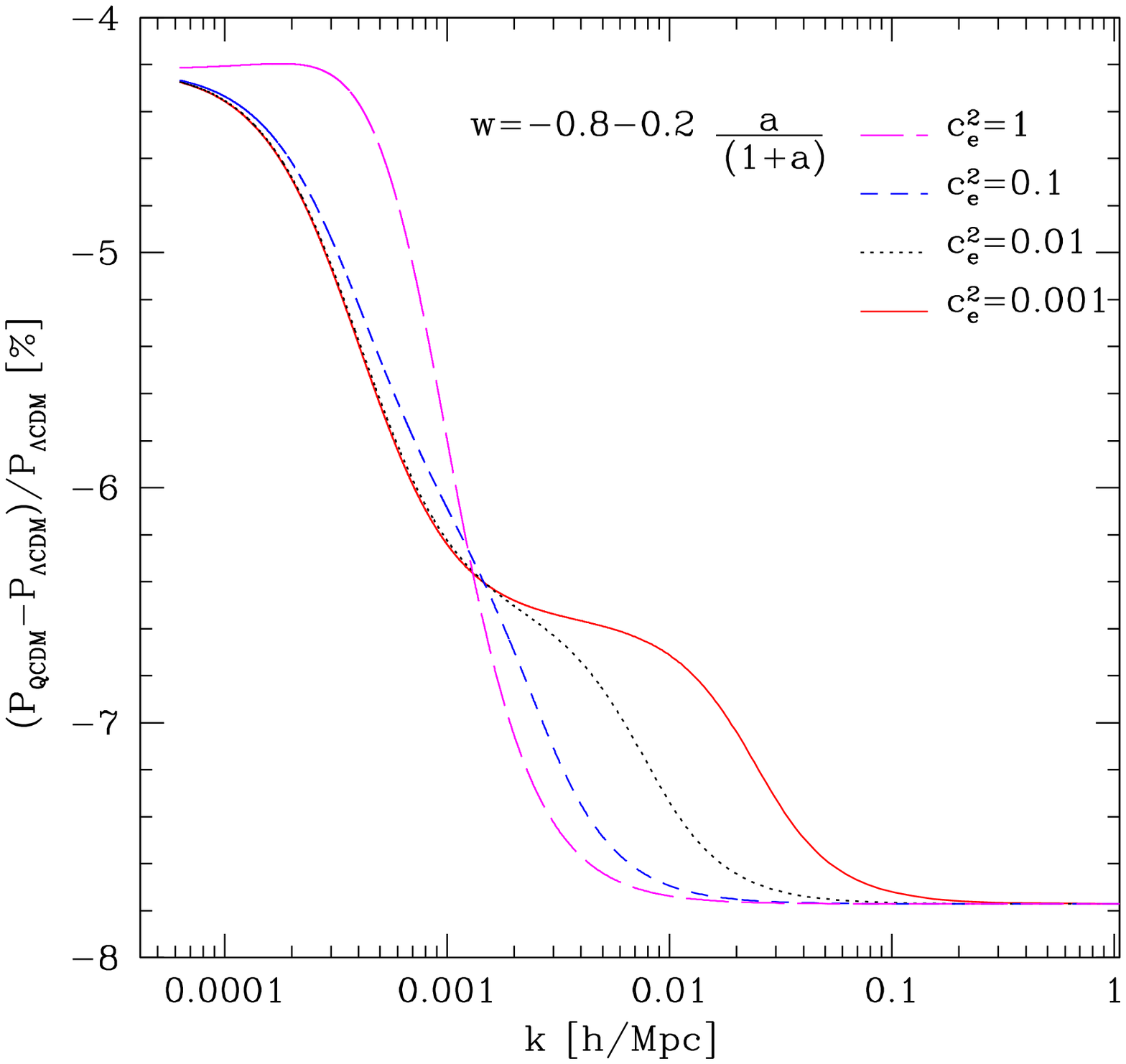}
\caption{Figure
showing suppression in matter power spectrum compared to $\Lambda$CDM model for the case when $w$ is evolving, but $c_e^2$ is constant.
In the left panel, $w$ is initially $-1$ and asymptotically approaches $-0.8$ whereas in the right panel $w$ is initially $-0.8$ and asymptotically approaches $-1$.
It can also be verified that in both the panel $w$ at the present epoch is $-0.9$.} \label{F3}
\end{figure}

\subsection*{Case II: Constant $c_e^2$ and evolving $w$}
When $c_e^2$ is constant, it is in general possible that the equation of state parameter is epoch dependent.
We will now consider such a case.
As mentioned earlier, for models with $F(X)=X^\alpha$,
$c_e^2$ is always a constant.
In fact, it is the form of the
potential $V(\phi)$ in these models which determines the evolution
of the EOS parameter. We consider the following parametrization
for the evolution of $w(a)$,
\begin{equation}\label{eqn of state paramaterization}
  w(a)=w_0+w_1 \l( \frac{a}{1+a}\r)^n
\end{equation}
where $w_0$, $w_1$ and $n$ are constants. The constant $w_0$
correspond to the initial value of the equation of state
parameter (at $a=0$). Asymptotically the equation of state
parameter approaches the value $w_0+w_1$. The parameter `$n$'
describes the rate at which $w$ evolves from its initial value $w_0$ to its asymptotic value $w_0+w_1$.
The value of $w(a)$ at the present epoch would be $w_0 + (w_1/2^{n})$.
Hence, greater the value of $n$ slower will the evolution $w$ from $w_0$ to its asymptotic value $w_0+w_1$.

The parametrization Eq.(\ref{eqn of state paramaterization}) is different from the one  generally investigated in the literature, viz.\ the Chevallier-Polarski-Linder (CPL) parametrization \cite{Chevallier-2001,Linder-2003} where $w(z)=w_0+w_{_{a}} \l( \frac{z}{1+z}\r)$.
However, at low red shifts the functional form of both the parametrization converges.
Although, the CPL parametrization is very well valid for the range of red shifts which is observationally relevant,
it does not restrict the value of $w$ to less than unity.
In fact in CPL parametrization, the equation of state parameter diverges asymptotically.
This problem does not arise in the parametrization of $w(a)$ introduced in Eq.(\ref{eqn of state paramaterization}).

\begin{figure}[t]
\includegraphics[scale=0.40]{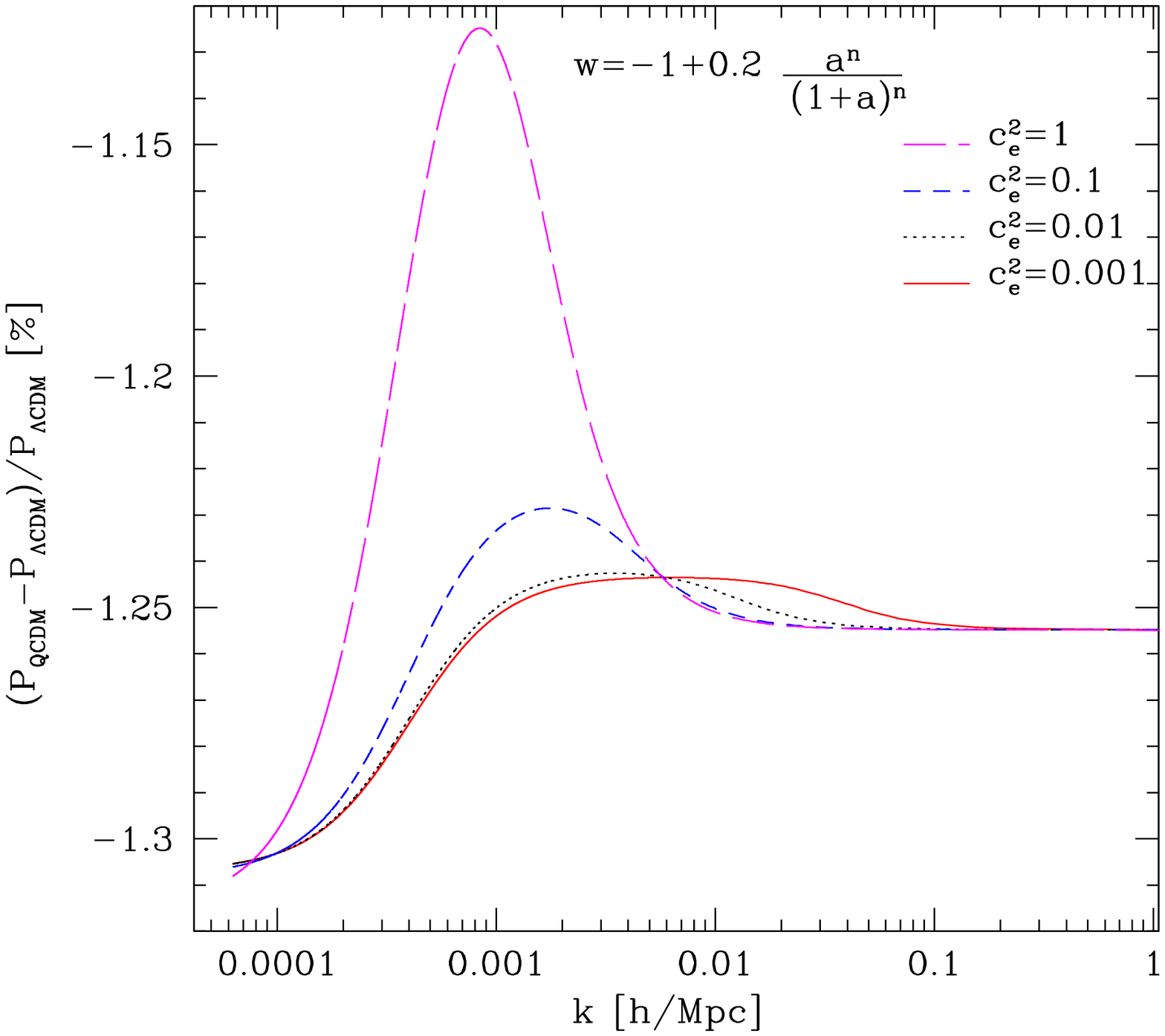}
\includegraphics[scale=0.40]{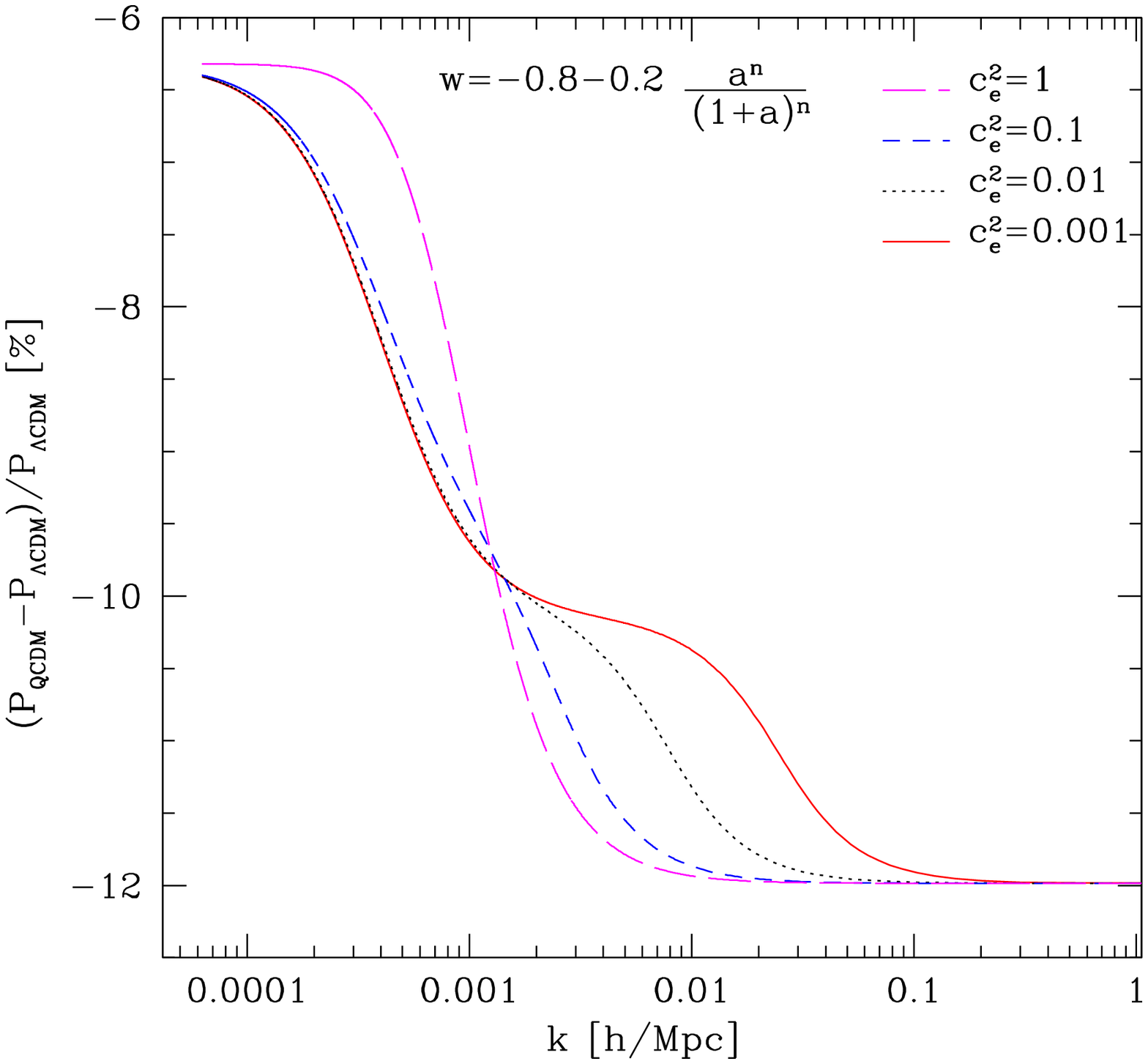}
\caption{Figure
showing suppression in matter power spectrum for  $w(a)$ given by Eq.(\ref{eqn of state paramaterization}) with $n = 10$.
In the left panel the equation of state parameter is initially $-1$ and stays close to it most of the time whereas in the right panel it evolves from  $-0.8$.
If one compares the suppression at two length scales say at $k = 0.1h/Mpc$ and at $k = 0.0001h/Mpc$, it
turns out that suppression is more at larger scale compared to small scales in the left panel of this figure.
The situation is exactly reverse in the plot on the right panel.
This illustrates the fact that the behavior of the suppression at these scales precisely depends on the functional form of $w(a)$.
} \label{F4}
\end{figure}

For constant value of $c_e^2$, we assume that $w(a)$ evolves with scale factor with the functional form given in Eq.(\ref{eqn of state paramaterization}).
In order to study the effect on matter power spectrum  by the dark energy
perturbation, we consider two specific behaviors of $w(a)$: one
in which $w(a)$ is initially -1 and approaches a constant
value ($w >-1$) asymptotically. The other case of interest is where
$w >-1$ initially, but asymptotically approaches -1.
For the case
where $w$ is initially -1 we choose $w_0 = -1$ and  $w_1$
to be 0.2, such that asymptotic value of $w=-0.8$. The range of
$w$ between -1 and -0.8 is well within the observable constraint
on the value of $w$. The second case of interest where $w$
decreases with time and finally approaches -1 would require
$w_0+w_1=-1$. We assume that initially $w=-0.8$, therefore $w_1$
would be  -0.2.

The suppression of CDM power spectrum as compared to the
$\Lambda$CDM in these two cases is shown in Fig.\ref{F3} for value $n = 1$ in Eq.(\ref{eqn of state paramaterization}). As in
the case of constant of $w$ and $c_e^2$, here also we see that
there is a suppression at all scales. The suppression is more in
the case when $w=-0.8$ initially (around 8 percent) than the case
where $w$ is initially -1. Again it can be seen from  Fig.
\ref{F3} the effect of different values $c_e^2$ on the suppression is
more pronounced at intermediate values of comoving wavenumber $k$.

 In the above cases we have chosen the parameter $n$ in Eq.(\ref{eqn of state paramaterization}) to be unity.
 If we increase $n$ to
 10, then the rate  at which EOS parameter switches
 its value from initial value to the asymptotic value $w_0+w_1$
 slows down dramatically. This case is shown in the Fig.\ref{F4}.
 When $n = 10$, the equation of state parameter for most the time stays near its initial value.
 Therefore, if $w_{0} = -1$, then it nearly mimic $\Lambda$CDM model for most of evolution, but deviates from it at the present epoch.
 The peculiar behavior of the suppression in the CDM power spectrum  in this case is shown in the left panel in Fig.\ref{F4}.
 Also we note from this plot that the suppression is very
 minimal in this case and contrary to other cases the suppression
 is maximum at small $k$'s than large values of $k$.
 This peculiar behavior is due to the choice of the function $w(a)$.

The right panel of Fig.\ref{F4} shows the other case
where $w$ is initially -0.8 and approaches -1, the percentage suppression in this case
 is high for the obvious reason that $w$ for most of the time stays close to $-0.8$.
 Also it should be  noted that suppression in this case is larger at smaller length scales unlike the case in the left panel.
 This illustrates  the fact  that the percentage suppression at both smaller and larger scales very much depends on the functional form of the equation of state parameter $w(a)$.
It should also be noted that the effect of different values of $c_e^2$ can  be seen more profoundly only at the intermediate scales.

\begin{figure}
\includegraphics[scale=0.40]{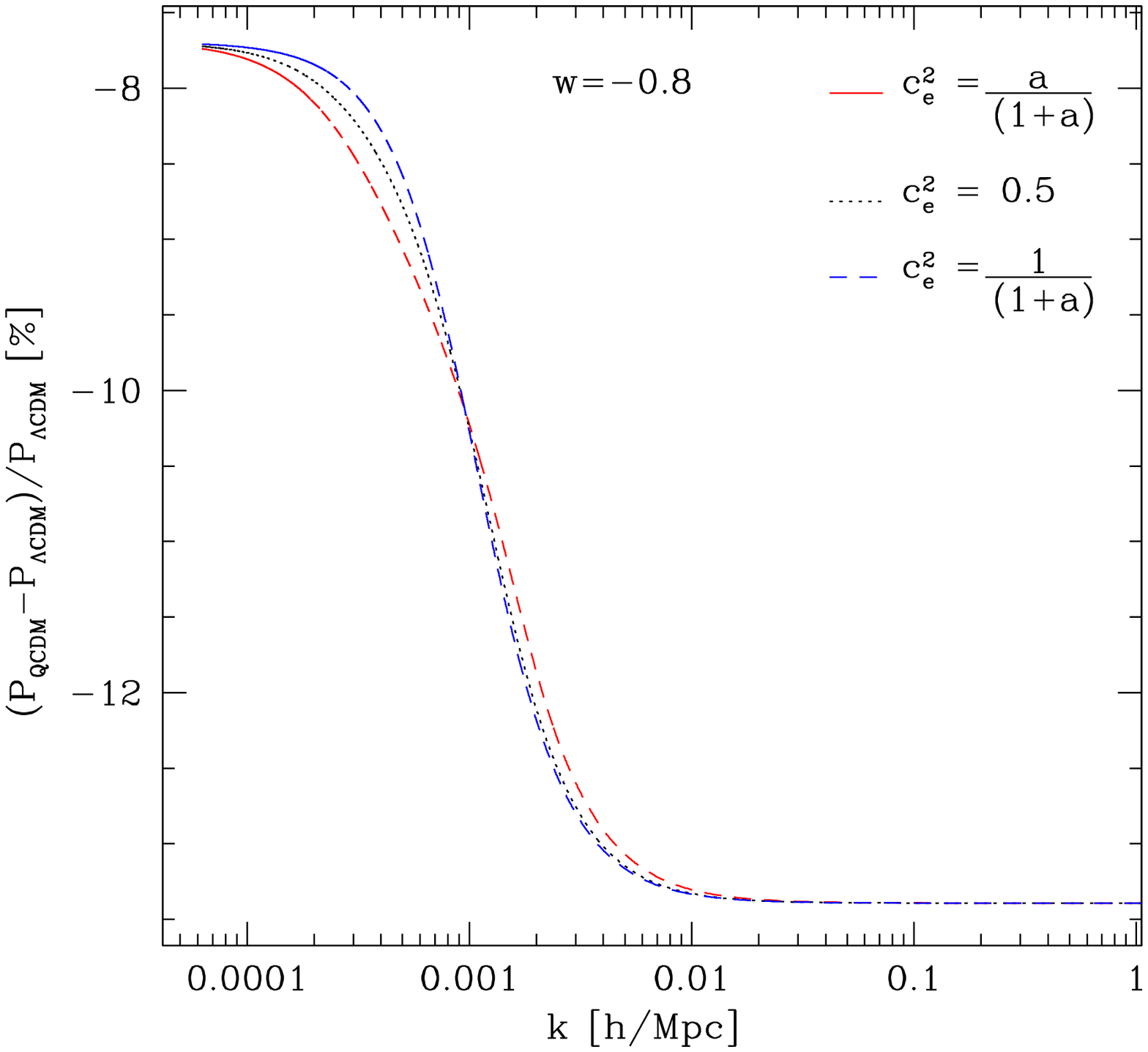}
\includegraphics[scale=0.40]{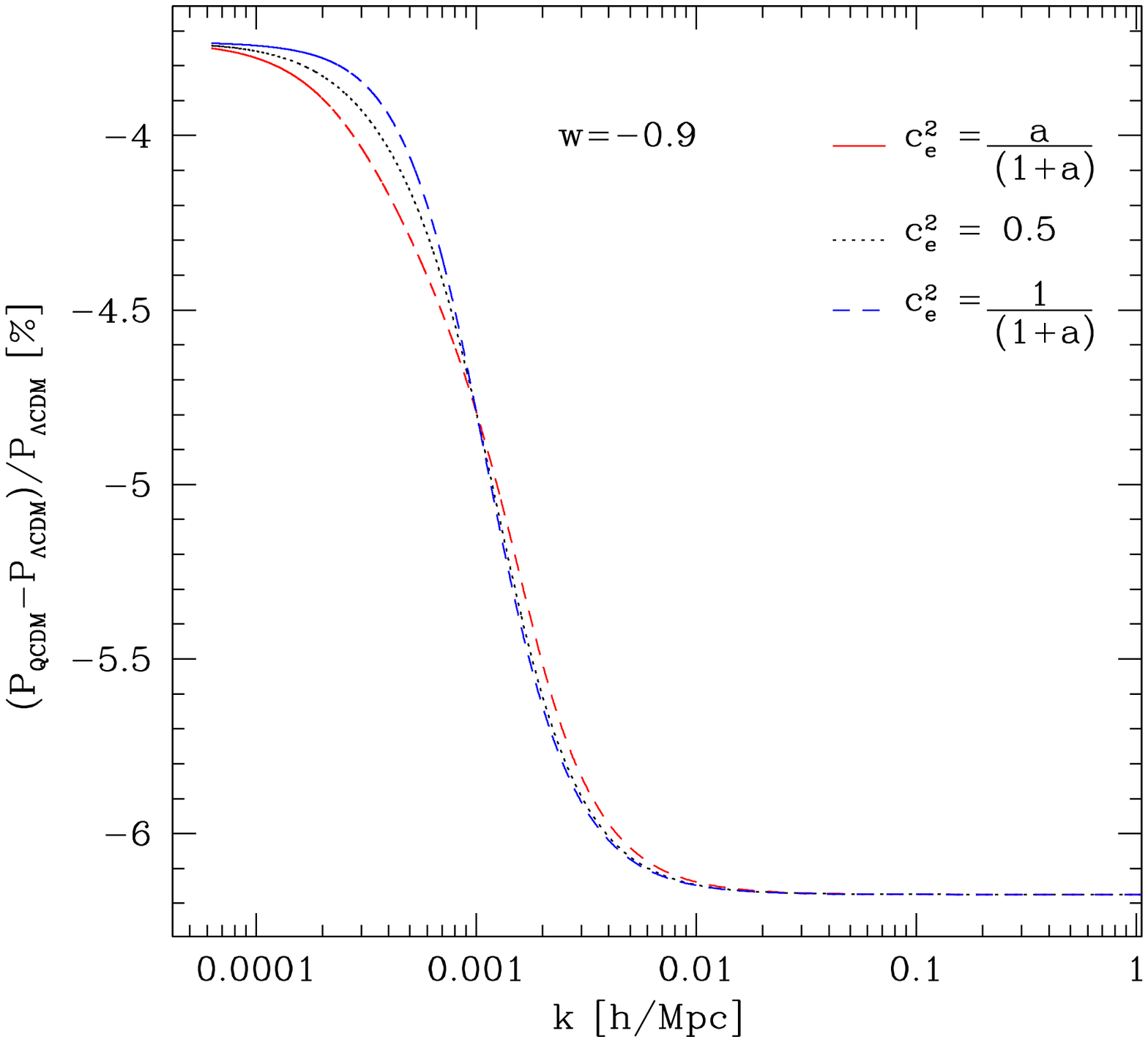}
\caption{The behavior of suppression in matter power spectrum as compared to $\Lambda$CDM model for the case when EOS parameter $w$ is constant but ESS parameter $c_{e}^{2}$ is evolving with scale factor is shown in this figure.
The left panel shows the case when $w = -0.8$ whereas the right panel show for $w = -0.9$.
The suppression in  matter power spectrum for the case when $c_{e}^{2}$ asymptotically decays to zero from its initial value of unity is compared with the reverse case where it increases to unity from zero.
For comparison $c_{e}^{2} = 0.5$ is also plotted.
Note that in all these cases the value of $c_{e}^{2}$ at the present epoch is $0.5$}\label{F5}
\end{figure}

\subsection*{Case III: constant $w$ and evolving $c_e^2$}
Let us now consider the case of dark energy models for which EOS parameter $w$ is constant but $c_e^2$
evolves with time.
As mentioned earlier, the class of dark energy models described by the
lagrangian density (\ref{Recons lagrang }) does allow $c_e^2$ to
evolve with time when $w$ is constant. It should be noted that
this is not in general valid for all classes of scalar field DE
models. For example in tachyon model of DE with $ \mathcal{L}_{1}
= V(\phi)\sqrt{1-2X}$, a constant $w$ always implies that $c_e^2$
is also constant.

We consider the following parametrization for
the evolution of $c_e^2$ with the scale factor
\begin{equation}
  c_e^2= c_0 + c_1 \l( \frac{a}{1+a}\r)^n\label{speed of sound paramaterization}
\end{equation}
This is similar to the parametrization of $w$ discussed in the
preceding section (see Eq.(\ref{eqn of state paramaterization})).
The interpretation of the constants $c_{0}$ and $c_1$ in Eq.(\ref{speed of sound paramaterization})
is as follows.
The constant $c_{0}$ corresponds to the initial value of the ESS parameter $c_e^2$,
whereas $c_{0} + c_1$ corresponds to its asymptotic value.
The parameter $n$ determines the rate at which the ESS parameter evolves from $c_0$ to its asymptotic value $c_0 + c_1$.

We are interested in comparing the suppression in CDM power spectrum with respect to $\Lambda$CDM  in cases where
$c_e^2$ decreases with time with those where it increases.
For the increasing case we choose $c_0=0$ and $c_1=1$ so that $c_e^2$ evolves from $0$ to 1 asymptotically.
 Similarly for the decreases  case we choose $c_0=1$ and $c_1 = -1$ so that $c_e^2$ evolves from $1$ to $0$.
The suppression of CDM power spectrum for these cases with n = 1 in Eq.(\ref{speed of sound paramaterization}) is shown in Fig.\ref{F5}.
This choice of these set of parameters ensures that the value $c_{e}^{2}$ at the present epoch  for these two cases is the same and is given by $0.5$.
Hence, for comparison the case of constant $c_{e}^{2}$ with $c_{e}^{2} = 0.5$ is also plotted in the same figure.
It is interesting to note that the three curves in Fig.\ref{F5} (both left and the right panel) almost overlap with each other indicating the fact that the effect of different  evolution of $c_{e}^{2}(a)$ for a constant $w$ is insignificant on the suppression of CDM power spectrum.
This may not be true for any evolution of $c_{e}^{2}$ with the scale factor.

\begin{figure}
\includegraphics[scale=0.40]{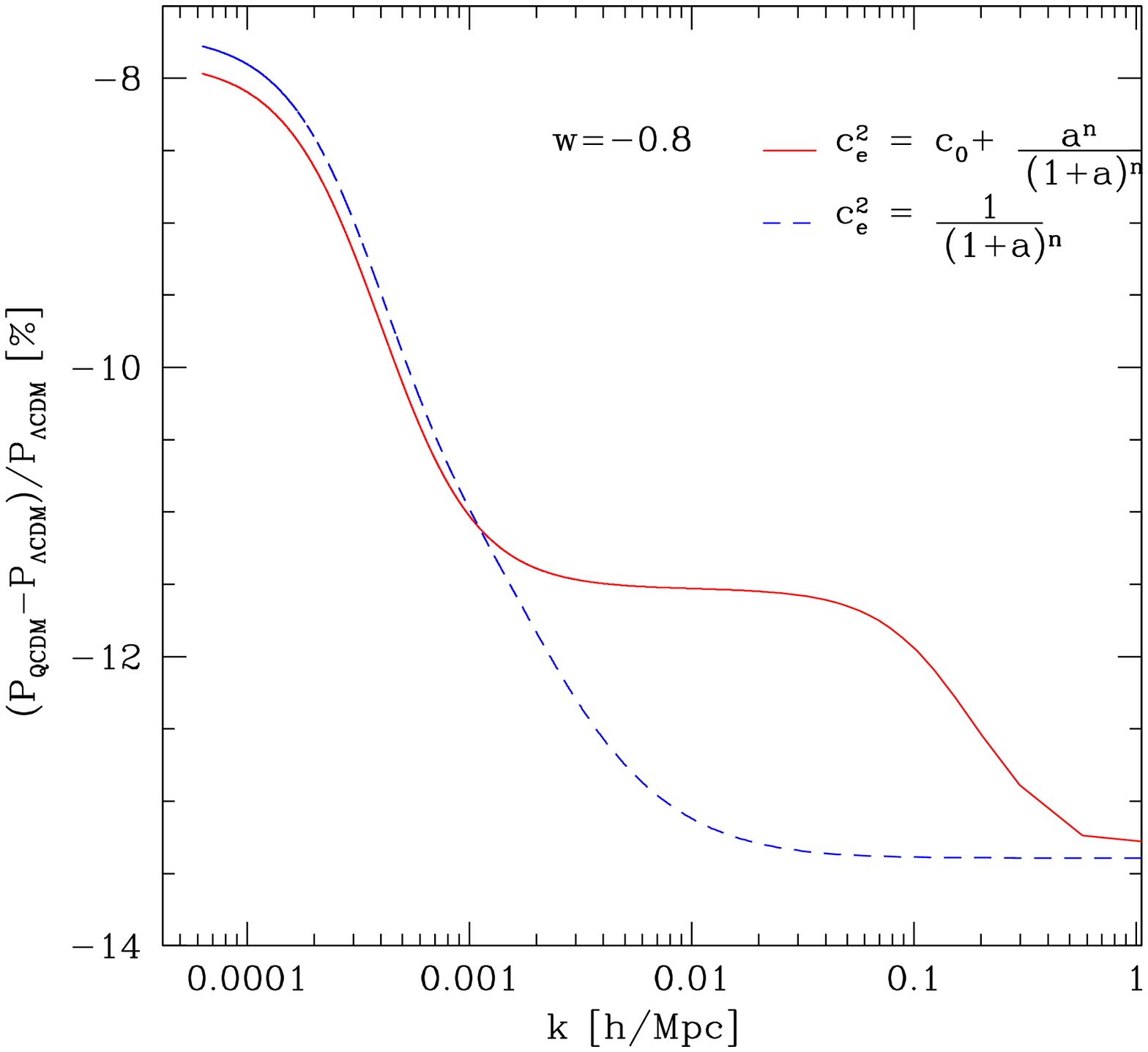}
\includegraphics[scale=0.40]{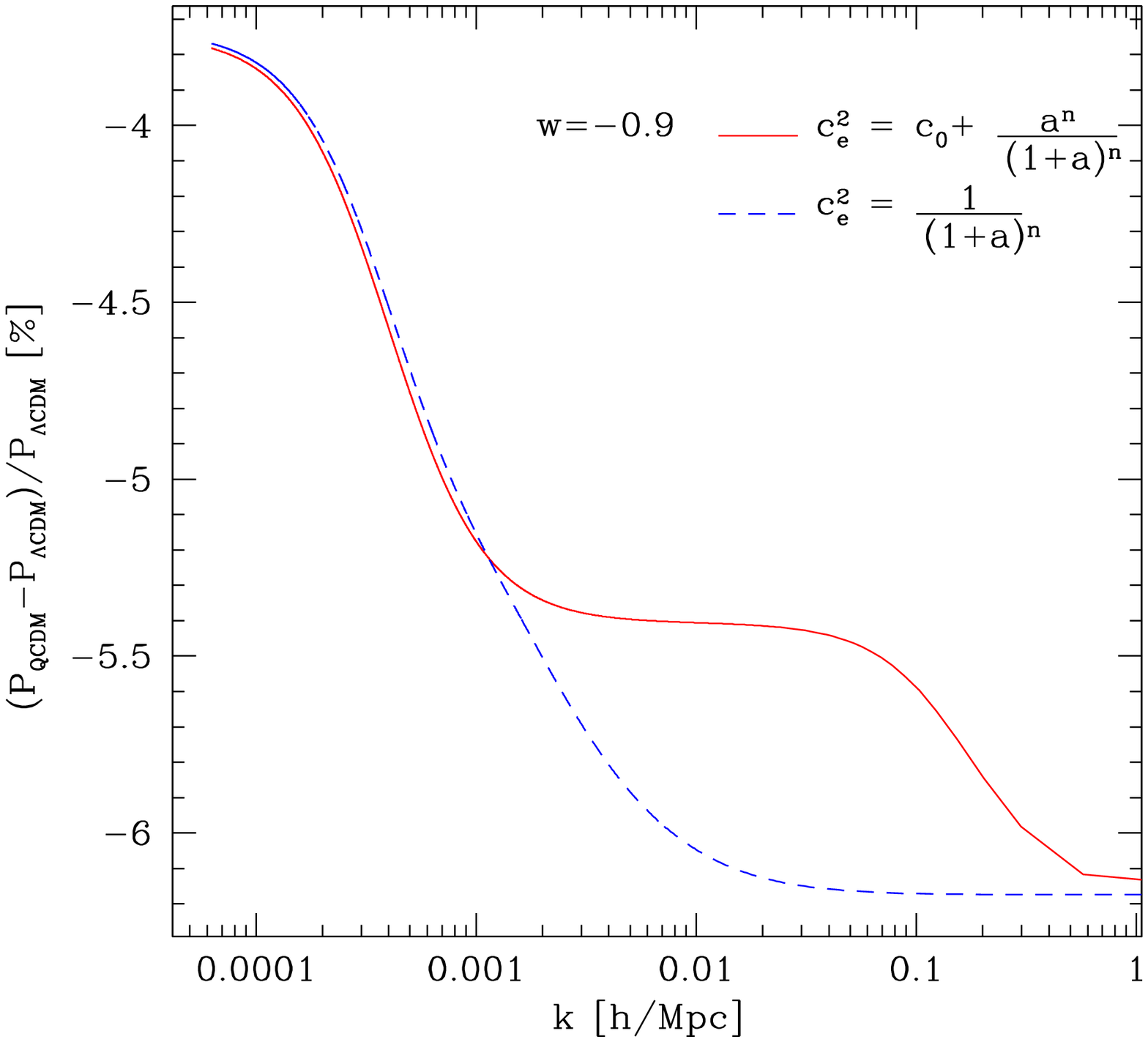}
 \caption{Comparison of suppression of
 matter power spectrum for two different evolution of   $c^2_e(a)$ with scale factor is shown in this Fig.
 Here we have taken $c_0=10^{-5}$ and $n=10$.
 For both these curves it can be verified that the present value of $c^2_e$ is approximately $10^{-3}$.} \label{F6}
\end{figure}

In order to check the generality  of this result we consider two more different evolution  of $c_{e}^{2}$.
The first one is the same function introduced in Eq.(\ref{speed of sound paramaterization}) with the following values of the parameter $c_{0} = 10^{-5}$, $c_{1} = 1$ and $n = 10$.
In this case $c_{e}^{2}(a)$ evolves from $c_{0} = 10^{-5}$ initially to $10^{-3}$ at the present epoch.
The second functional form of $c_{e}^{2}(a)$ we consider is $c_{e}^{2}(a) = 1/(1 + a)^{n}$ with $n = 10$.
Here, in this case $c_{e}^{2}(a)$ evolves from its initial value of unity to $10^{-3}$ at the present epoch.
The suppression of CDM power spectrum in these two cases is shown in Fig.\ref{F6}
It is evident from this figure that the two curves are not as close to each other as compared to the three curves in Fig.\ref{F5}.
The reason for this is that in these two curves order of magnitude change in  $c_{e}^{2}$ from $z = 1000$ to present epoch is much larger than the cases presented in Fig.\ref{F5}.
However, it should be noted that, as in the previous cases of constant
 $c_e^2$ the effect of different choice of  $c_e^2(a)$ on CDM power spectrum is not evident at large and low values of $k$.

\subsection*{Case IV: Evolving $w$ and  $c_e^2$}
Until now we have considered the cases where both $w$ and $c_e^2$ are constants and the case where either one
of them is epoch dependent.
For the purpose of completeness, we now consider the case where both $w$ and $c_e^2$ evolves
with time.
This is the most general case and in reality, for k-essence models, it is likely that these two parameters are epoch dependent.

\begin{figure}
{\includegraphics[scale=0.40]{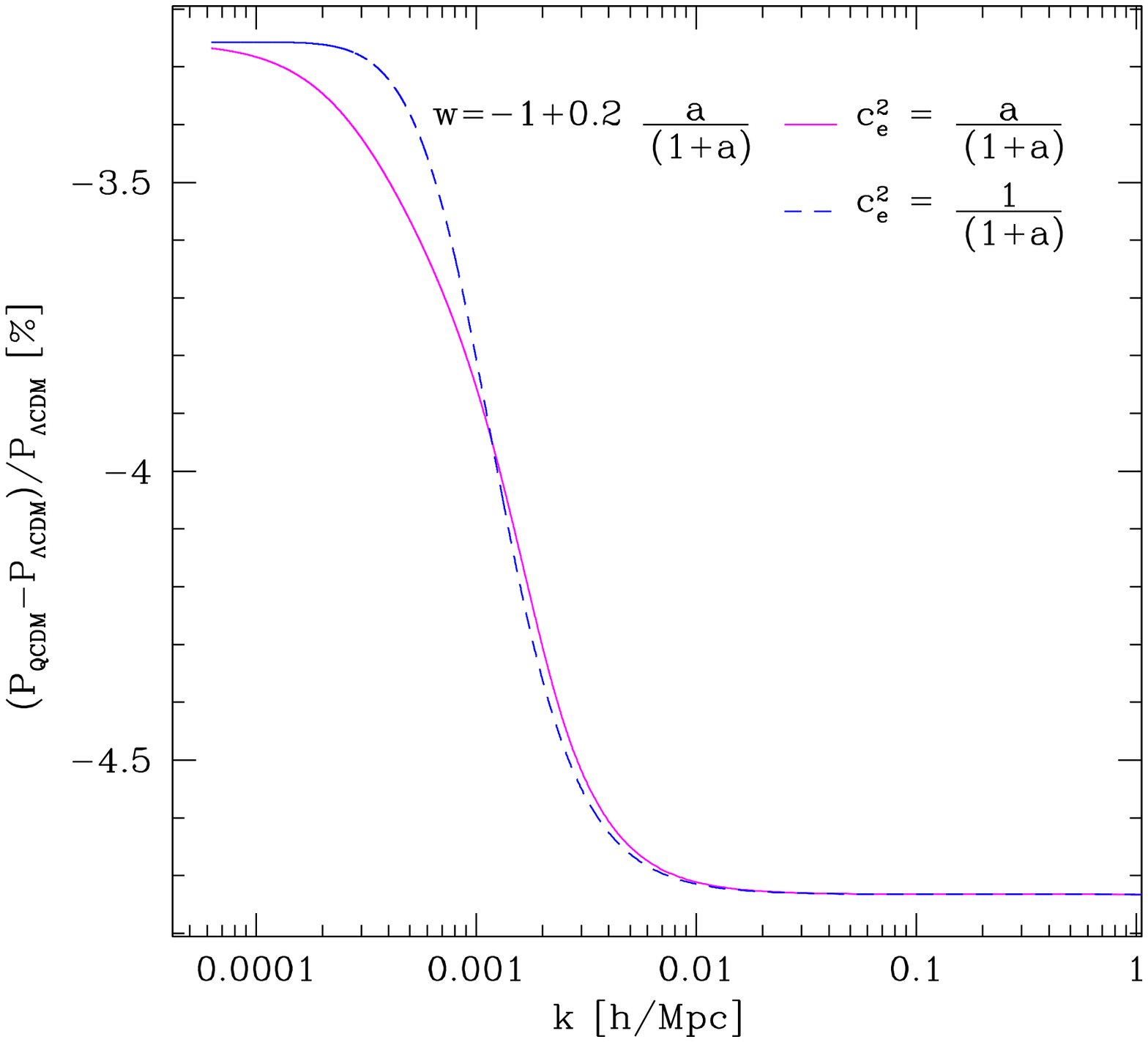}a}
{\includegraphics[scale=0.40]{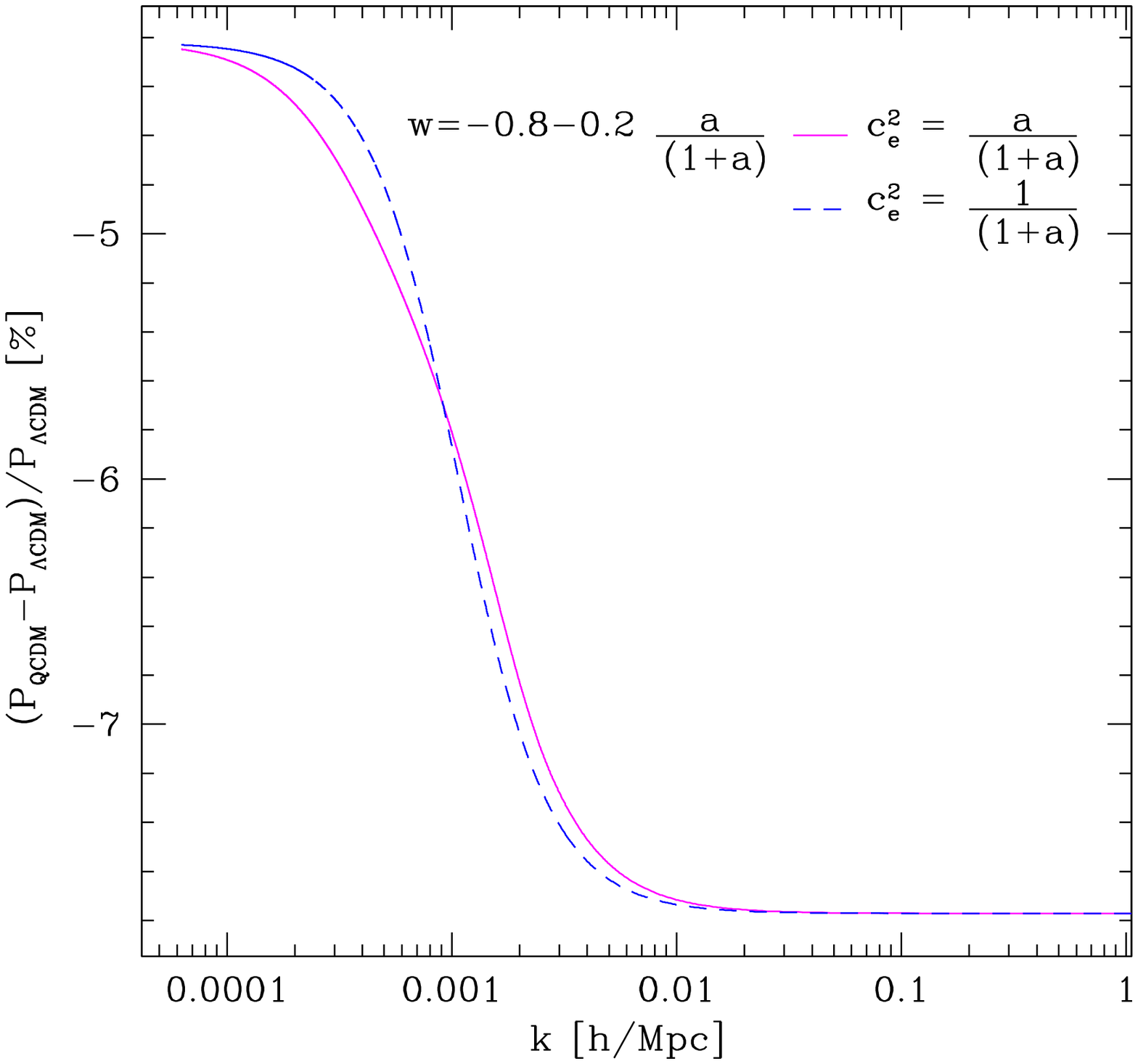}b} \caption{Supression
in matter power spectrum as compared to that in $\Lambda$CDM model in the case when both  $c_e^2$ and $w$ are epoch dependent.
In the
left panel we have chosen  a form of $w(a)$, which is initially $-1$ and
asymptotically goes to a constant value $-0.8$ and in the right panel $w$
is chosen such that it is initially $-0.8$ and asymptotically approaches $-1$.}\label{F7}
\end{figure}

We will study two cases, first with the following
evolution of $w(a)= -1+0.2\l( \frac{a}{1+a}\r)$,  and the two
functions $c_e^2= \l( \frac{a}{1+a}\r)$ and $c_e^2= \l(
\frac{1}{1+a}\r)$.
The choice of $w(a)$ corresponds to the case where it evolves from $-1$ to $-0.9$ at the present epoch
The second case we consider is  with the same two functional form of
$c_e^2$ but with $w(a)$ given by  $w(a)= -0.8-0.2\l( \frac{a}{1+a}\r)$ is initially -0.8
and evolves to $-0.9$ at the present epoch.
The suppression of CDM power spectrum as compared to $\Lambda$CDM model in these two cases are shown
in the Fig.\ref{F7}.
it is evident from this figure that the behavior of suppression with $k$
is similar in the both cases, with suppression being maximum at
the smaller scales  than at larger scales.
Also it is clear from the
two figures that suppression is more in the case where $w$ is
initially $-0.8$ than the case where initial value is -1,
reiterating the point that as we move away from value of $w=-1$
the the effect of dark energy perturbations is more important.
Finally, we
infer that for different functional form of  $c_e^2(a)$ for a given evolution of $w(a)$  does not have any considerable
effect on the matter power spectrum.
In fact it can be observed from Fig.\ref{F7} that changing the  functional form of $w(a)$ for a given evolution of
$c_e^2(a)$ influences the matter power spectrum more severely than the reverse case.

\section{The role of perturbation in dark energy}
From the discussions in the preceding section, it follows that for a given evolution $w(a)$, the effect of different functional form of $c_{e}^{2}$ on the CDM power spectrum is significant only at the intermediate scales (see Fig.\ref{F1} to Fig.\ref{F7}).
In fact, at scales much smaller (or larger) than the Hubble radius, the suppression of CDM power spectrum with respect to the $\Lambda$CDM model is nearly independent of the functional form of  $c_{e}^{2}(a)$.
The reason for this behavior is discussed in this section.

\subsection*{At the sub-Hubble scales}
\begin{figure}
{\includegraphics[scale=0.40]{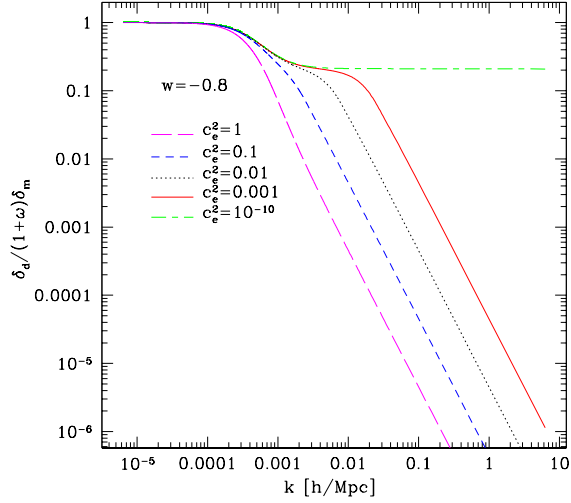}} \caption{The raio of perturbation in dark energy to the perturbation in dark matter scaled by a factor $1/(1 + w)$ is plotted in this figure.
It can be observed that at length scales much larger than the Hubble radius  this ratio approaches unity.
The reason for this is that at these scales the incurvature perturbation vanishes.
Therefore, at these scales $\delta_{d} = (1 +w)\delta_{m}$.}\label{F8}
\end{figure}

At scales much smaller than the Hubble radius, the perturbation in dark energy is negligibly smaller than the corresponding perturbation in dark matter \cite{Sanil-2008a}.
This result follows from the numerical solutions of Eqs.(\ref{Phi prime}) to (\ref{delta d}).
In Fig.\ref{F8}, we have plotted the ratio $\delta_{d}/(1 +w)\delta_{m}$ at the present epoch for different length scales of perturbations.
It follows from this figure that at scales much smaller than the Hubble radius ($k >> 0.001h/Mpc$), the perturbation in dark energy is negligibly smaller  in comparison to that in dark matter.
In fact, at these scales it can be assumed that the dark energy is homogeneously distributed.

Since we have assumed dark energy to be a scalar field, homogeneous dark energy at these scales reflects to the fact that scalar field fluctuations at these scales are negligibly smaller than other perturbation variables such as $\delta_{m}$, $\Phi$, etc.
With this assumption that the dark energy is homogeneous at scales much smaller than the Hubble radius,
it turns out that the evolution of the Bardeen potential $\Phi$ at these scales is governed by the following equation
\begin{equation}
\ddot{\Phi} + 4H\dot{\Phi} + \left(2\dot{H} + 3H^{2} + 3H^{2}\l(\frac{1 + w(a)}{2}\r)\Omega_{d}(a)\right)\Phi
= 0. \label{phi eqn at sub hubble}
\end{equation}

The above equation corresponds to the evolution of $\Phi$ in a system of dark matter and homogeneous scalar field dark energy.
In the case of $\Lambda$CDM  model, the corresponding equation for $\Phi$ is given by
\begin{equation}
\ddot{\Phi}_{\Lambda} + 4H\dot{\Phi}_{\Lambda} + \left(2\dot{H} + 3H^{2}\right)\Phi_{\Lambda}
= 0. \label{phi eqn LCDM}
\end{equation}
This evolution  equation for $\Phi$ in $\Lambda$CDM  model is valid at all length scales.
However, it should be noted that Eq.(\ref{phi eqn at sub hubble}) is strictly valid only at scales much smaller than
the Hubble radius where dark energy can be assumed to be homogeneous.
At these scales, the perturbation in dark matter is given by
\begin{equation}
\delta_m = -\l(\frac{1}{4 \pi G\,\bar{\rho}_m(a)}\r)\frac{k^{2}}{a^{2}}\Phi\label{deltam sun hubble}
\end{equation}
Therefore, the percentage suppression in CDM power spectrum due to dark energy in comparison to $\Lambda$CDM model
($\Delta \%$ defined in Eq.(\ref{Delta})) would become
\begin{equation}\label{Delta1}
\Delta \%  =  \frac{\Phi^{2} - \Phi_{\Lambda}^{2}}{\Phi_{\Lambda}^{2}}\times 100
\end{equation}

From the numerical solution of Eqs.(\ref{phi eqn at sub hubble}) and (\ref{phi eqn LCDM}),
it follows that for $w = -0.8$, the value of $\Delta \%$ is $-13.39$ whereas for $w = -0.9$, it turns out to be $-6.17$.
These values matches with the one shown in Fig.\ref{F1} at $k >> 0.001 h/Mpc$, which was plotted without assuming a priori  the homogeneity of  dark energy at these scales.

The evolution of Bardeen potential $\Phi$  at scales much smaller than the Hubble radius is independent of $c_{e}^{2}$ (see Eq.(\ref{phi eqn at sub hubble})).
The corresponding growth of CDM perturbation at these scales given by Eq.(\ref{deltam sun hubble}) is also independent of $c_{e}^{2}$.
Therefore, at these scales the suppression in CDM power spectrum is independent of the dark energy speed of sound $c_{e}^{2}$.
As emphasized earlier, the primary reason for this is the fact that at these scales dark energy is nearly homogeneous.
It is only through the perturbation in dark energy that $c_{e}^{2}$ plays a role in suppressing the CDM power spectrum compared to that in $\Lambda$CDM model.

\subsection*{At the super-Hubble scales}

At scales comparable to Hubble radius and beyond ($k < 0.001h/Mpc$), the perturbation in dark energy cannot, in general, be neglected in comparison to that in dark matter.
This evident from Fig.\ref{F8}, which shows that  in the limit $k << 0.001h/Mpc$, the perturbation in dark energy becomes
\begin{equation}
\delta_{d} = (1 + w)\delta_{m}\label{delta d by delta m}
\end{equation}

If $w \neq -1$, then perturbation in dark energy can become comparable to that in matter at these scales ($k < 0.001h/Mpc$).
Eq.(\ref{delta d by delta m}) also follows from  the fact that at these scales,
the isocurvature perturbation or the total non-adiabatic pressure perturbation vanishes.
In addition to this, the intrinsic entropy perturbation of dark energy also vanishes at these scales.
The intrinsic entropy perturbation of dark energy  is proportional to its non-adiabatic pressure perturbation defined as
\begin{equation}
\delta p_{_{nad}} = \delta p_{d} - c_{a}^{2}\delta \rho_{d}
\end{equation}
For scalar field dark energy, it $\delta p_{_{nad}}$ can also be expressed as
\begin{equation}
\delta p_{_{nad}} = (c_{e}^{2} - c_{a}^{2})\l[\delta_{d} - 3H^{2}(1 + w)a^{2}u_{d}\r]\bar{\rho}_{d}(a).
\end{equation}
At the super Hubble scales, the $\delta p_{_{nad}}$ of dark energy vanishes not because of the fact that $c_{e}^{2} = c_{a}^{2}$ but because
\begin{equation}
\delta_{d} = 3H^{2}(1 + w)a^{2}u_{d}\label{delta d and um}
\end{equation}
In fact, it can be verified that in the limit $k \rightarrow 0$, Eqs.(\ref{delta d by delta m}) and (\ref{delta d and um}) are consistent with the covariant conservation equation for dark matter and dark energy.

With the fact that the non-adiabatic pressure perturbation of dark energy vanishes at scales much larger than the Hubble radius, it turns out that the evolution of Bardeen potential at these scales is governed by the following equation
\begin{equation}
\ddot{\Phi} + \l(4 + 3c_{a}^{2}(1 + w)\l(\frac{\Omega_{d}(a)}{1 + w(a)\Omega_{d}(a)}\r)\r)H\dot{\Phi} + \left(2\dot{H} + 3H^{2} + 3H^{2}c_{a}^{2}(1 + w)\l(\frac{\Omega_{d}(a)}{1 + w(a)\Omega_{d}(a)}\r)
\right)\Phi
= 0. \label{phi eqn at super hubble}
\end{equation}
The evolution of $\Phi$ in $\Lambda$CDM model follows from the above equation if we substitute $w = -1$.
In the limit $k << 0.001h/Mpc$, neglecting the $k^{2}$ term in time-time linearized Einstein's equation ($\delta G^{0}_{ \hspace{0.2cm} 0} = 8\pi G\,\delta T^{0}_{\hspace{0.2cm} 0}$,) together with the fact that $\delta_{d} = (1 + w)\delta_{m}$, it turns out that the equation for $\delta_{m}$ becomes
\begin{equation}
\delta_{m} = \l(\frac{-2}{1 + w(a)\Omega_{d}(a)}\r)\l(\Phi + \frac{\dot{\Phi}}{H}\r)
 \label{delta m at super hubble}
\end{equation}

The percentage suppression in CDM power spectrum can now be evaluated by numerically solving Eq.(\ref{phi eqn at super hubble}) and then evaluating $\delta_{m}$ using Eq.(\ref{delta m at super hubble}).
It turns out $\Delta \%$ defined in Eq.(\ref{Delta}) for $w = -0.8$ at these scales ($k \rightarrow 0$) is $-7.7$.
and for $w = -0.9$  it becomes $-3.7$.
These values are consistent with those in Fig.\ref{F1} for $k << 0.001h/Mpc$ where we have solved the exact perturbation equation without imposing any assumption.

Hence, at scales much higher than the Hubble radius, the effective speed of sound of dark energy does not influence the CDM perturbations.
This is  because the intrinsic entropy perturbation of dark energy vanishes at these scales.
It is the intrinsic entropy perturbation of dark energy which carries the information of the ESS parameter $c_{e}^{2}(a)$.
Thus, it is only at the intermediate scales (scales around $k \sim 0.01h/Mpc$) where the effect of different evolution of $c_{e}^{2}(a)$ is more pronounced on the suppression of CDM power spectrum with respect to $\Lambda$CDM model.

\section{Summary and Conclusions}
In this paper we have investigated the influence of perturbation in dark energy
 on the matter power spectrum in models where the effective speed of sound $c_{e}^{2}(a)$ of dark energy evolves with time.
We first presented a method of reconstructing the lagrangian density of dark energy of the form $\mathcal{L}(X,\phi) = F(X)-V(\phi)$ from a given evolution of $c_{e}^{2}(a)$ and the equation of state parameter $w(a)$.
This illustrates the fact that these scalar field dark energy models, in principle, allows a
wide class of solutions with different functional form of $w(a)$ and  $c_{e}^{2}(a)$.
In fact, in these models evolution in $c_{e}^{2}(a)$ can be independent of that in $w(a)$.

We then investigated the growth of CDM perturbation as influenced by the perturbations in dark energy in models where $(i)$ both $c_{e}^{2}$ and $w$ are constants,
 $(ii)$ either of them are epoch dependent and
 $(iii)$ both of them are epoch dependent.
 It is shown that in all these cases, the CDM power spectrum is generically suppressed in comparison to that in $\Lambda$CDM model.
 The degree of suppression at different length scales does in fact depend on the behavior of $w(a)$ and  $c_{e}^{2}(a)$.

When the equation of state parameter of dark energy is constant, it is shown that the percentage suppression in  CDM power spectrum with respect to the $\Lambda$CDM model decreases with increasing length scale of perturbation (see Fig.\ref{F1}).
However, this is not generically true for any evolution of $w(a)$.
In fact, it is found that for slow evolution of $w(a)$ from $-1$ to a value close to it (with $w + 1 > 0$),
the percentage suppression in  CDM power spectrum can increase with increase in the length scale of perturbation (see, left panel of Fig.\ref{F4}).

Primarily, in this paper we have compared the percentage suppression in  CDM power spectrum in dark energy models where its effective speed of sound $c_{e}^{2}(a)$  increases with scale factor with those where it decreases.
It is shown that the effect of different evolution of $c_{e}^{2}(a)$ of dark energy on the matter power spectrum
for a given evolution of equation of state parameter $w(a)$ is not as much significant as compared to the reverse case, viz.\ different evolution of $w(a)$ for a given evolution of $c_{e}^{2}(a)$ (as shown in Fig.\ref{F5} to Fig.\ref{F7}).
This illustrates the fact that the effect of equation of state parameter $w(a)$ of dark energy  on the CDM power spectrum is much more severe than its effective speed of sound $c_{e}^{2}(a)$.

Further, it is also shown that the effect of different evolution of $c_{e}^{2}(a)$ for a given evolution of $w(a)$, on the suppression of  CDM power spectrum is more pronounced only at the intermediate scales at around $k \sim 0.01h/Mpc$.
In fact it is observed that the degree of suppression of  CDM power spectrum with respect to $\Lambda$CDM model
is nearly independent of $c_{e}^{2}(a)$ at scales much smaller and larger than the Hubble radius.
The reason for this behavior at scales much smaller than  the Hubble radius is as follows.
At these scales, it turns out that the perturbation in dark energy is negligibly smaller than the corresponding perturbation in matter.
One can effectively approximate dark energy to be homogeneous at these scales.
Since it is the perturbation in dark energy which carries the information of its effective speed of sound,
the suppression of  CDM power spectrum at these scales will be independent of any functional form of $c_{e}^{2}(a)$.
The suppression at these scales will primarily be governed by the functional form of $w(a)$ of dark energy.

Even at scales much larger than the Hubble radius,  suppression of  CDM power spectrum with respect to $\Lambda$CDM model is nearly independent of $c_{e}^{2}(a)$.
At these scales the perturbation in dark energy cannot be neglected in comparison to matter perturbation if $w(a)$ deviates from $-1$ (as shown in Fig.\ref{F8}).
However, it is found that at these scales both the total entropy perturbation (or the isocurvature perturbation) and the intrinsic entropy perturbation of dark energy vanishes.
It is in fact the intrinsic entropy perturbation of dark energy which encodes the information of its effective speed of sound.
Therefore, at these scales, although there is dark energy perturbation, it is purely adiabatic and consequently
its effect on the suppression of  CDM power spectrum will be same for different $c_{e}^{2}(a)$.

In summary,  it is illustrated in this paper that the suppression of  CDM power spectrum with respect to $\Lambda$CDM model both at scales much larger and smaller than the Hubble radius only depends  on the form of $w(a)$ of dark energy.
It is only at the intermediate scales, a non zero value of the effective speed of sound of dark energy leaves its imprint on the CDM power spectrum.
Precisely determining the effective speed of sound of dark energy, although observationally challenging \cite{Putter-2010}, is necessary to understand its nature.
We hope that future observations will shed more light on the `\emph{darkness}' of the dark energy.

\section*{Acknowledgments}
We thank Nisha Katyal and Sowgat Muzahid for the help in preparation of few figures.

\end{document}